\documentclass{nature}

\usepackage{amsmath,amssymb,enumerate}
\usepackage{lineno}
\usepackage{graphicx,float,color}
\usepackage[font={small,singlespacing},labelfont=bf]{caption}
\DeclareCaptionLabelSeparator{vline}{~$\big\rvert$ }
\usepackage{txfonts} 
\usepackage{comment}
\usepackage[dvipsnames]{xcolor}
\usepackage[colorlinks,citecolor=blue,linkcolor=magenta]{hyperref}
\usepackage{textcomp}
\usepackage{siunitx}

\title{Double-network-inspired mechanical metamaterials}

\author{James Utama Surjadi$^{1}$, Bastien F. G. Aymon$^{1}$, Molly Carton$^{1}$, Carlos M.~Portela$^{1\ast}$}
\begin{document}
\maketitle

\begin{affiliations}
\footnotesize
 \item Department of Mechanical Engineering, Massachusetts Institute of Technology, Cambridge, MA 02139, USA
\end{affiliations}
\vspace{10pt}
\spacing{1.0}

\begin{abstract}
Mechanical metamaterials are renowned for their ability to achieve high stiffness and strength at low densities, often at the expense of low ductility and stretchability---a persistent trade-off in materials. In contrast, materials such as double-network hydrogels feature interpenetrating compliant and stiff polymer networks, and exhibit unprecedented combinations of high stiffness and stretchability, resulting in exceptional toughness. Here, we present double-network-inspired (DNI) metamaterials by integrating monolithic truss (stiff) and woven (compliant) components into a metamaterial architecture, which achieve a tenfold increase in stiffness and stretchability compared to their pure woven and truss counterparts, respectively. Nonlinear computational mechanics models elucidate that enhanced energy dissipation in these DNI metamaterials stems from increased frictional dissipation due to entanglements between the two networks. Through introduction of internal defects, which typically degrade mechanical properties, we demonstrate an opposite effect of a threefold increase in energy dissipation for these metamaterials via failure delocalization. This work opens avenues for developing new classes of metamaterials in a high-compliance regime inspired by polymer network topologies.

\end{abstract}
\newpage

\subsection{Main}
\hfill\\
Historically, the development of mechanical metamaterials has predominantly focused on attaining high stiffness and strength while maintaining low densities, with architectures ranging from classical periodic truss lattices to plate and shell lattices that can achieve stiffness and strength close to theoretical limits.\cite{bauer2017nanolattices, surjadi2019mechanical, greer2019three, schwaiger2019extreme} However, sharp joints and nodes typically induce stress concentration that compromises ductility or stretchability, especially in truss and plate lattices.\cite{meza2014strong, zheng2014ultralight, shaikeea2022toughness, berger2017mechanical, tancogne20183d, crook2020plate} On the contrary, recently reported stretchable metamaterials, exemplified by helical\cite{yan2020soft} and woven lattices,\cite{moestopo2020pushing, moestopo2023knots} significantly depart from the conventional paradigm. Nevertheless, the stiffness of these stretchable designs is often multiple orders-of-magnitude lower than traditional high-stiffness lattices. This pronounced stiffness-stretchability trade-off, which is also commonly observed in most existing material systems, fundamentally constrains the energy dissipation capabilities of metamaterials.

An example of a pivotal shift in this material-property trade-off was marked by hydrogels, specifically double-network (DN) hydrogels,\cite{gong2003double} which feature interpenetrating sparsely crosslinked (compliant) and densely crosslinked (stiff) polymer networks. These DN hydrogels can attain both high stiffness and stretchability, overcoming the longstanding stiffness-stretchability trade-off in conventional single-network hydrogels.\cite{nakayama2004high, sun2012highly, matsuda2019mechanoresponsive} The lack of strength and toughness observed in conventional, cross-linked single-network hydrogels can be attributed to the low density of polymer chains and limited entanglements within the chain structure.\cite {gong2010double, kim2021fracture} Conversely, in a DN hydrogel, the rupture of covalent bonds in the stiff network upon mechanical loading serves as sacrificial-bond breakage to dissipate energy. In addition, the compliant network prevents catastrophic failure via entanglements, inducing additional energy dissipation while enabling extensive fragmentation of the stiff network throughout the entire hydrogel.\cite{huang2007importance, yu2009direct, gong2010double} This synergistic interplay between the two networks gives rise to extraordinary toughness greater than the sum of the stiff and compliant counterparts independently.

Here, by drawing inspiration from DN hydrogels, we propose a route to design and fabricate double-network-inspired (DNI) metamaterials with extreme combinations of stiffness and stretchability, integrating both monolithic-truss and woven architectures to act as stiff and compliant networks, respectively. \emph{In situ} micromechanical tensile experiments on the metamaterials showcase both high stiffness and stretchability with ultimate stretches exceeding 4, providing an order-of-magnitude increase in ultimate deformation compared to their constituent materials. These mechanical responses significantly outperform those reported for other metamaterials,\cite{moestopo2020pushing, moestopo2023knots, zheng2016multiscale, jiang2021centimetre, mateos2019discrete, bauer2015push, montemayor2016insensitivity} resulting in a  multi-fold increase in specific energy dissipation compared previously reported metamaterial designs. By employing nonlinear finite element models (FEM), we establish that the enhanced energy dissipation can be linked to increased entanglements between the monolithic and woven components, augmenting frictional dissipation. Notably, the introduction of internally distributed defects in the stiff network further amplifies energy dissipation by delocalizing failure, a departure from the typical scenario where defects often result in the degradation of mechanical properties for amorphous materials.\cite{shaikeea2022toughness, glaesener2023predicting} Our work not only provides a strategy for mitigating the stiffness-stretchability trade-off in existing metamaterials but also charts a route for creating new classes of metamaterials inspired by polymer networks. This new design paradigm could unlock unprecedented combinations of mechanical/functional properties and potentially serve as analog systems to further our understanding on the behavior of complex polymer systems.

\subsection{Double-network-inspired design paradigm}
\begin{figure}[H]
    \centering
    \includegraphics[width=0.84\textwidth]{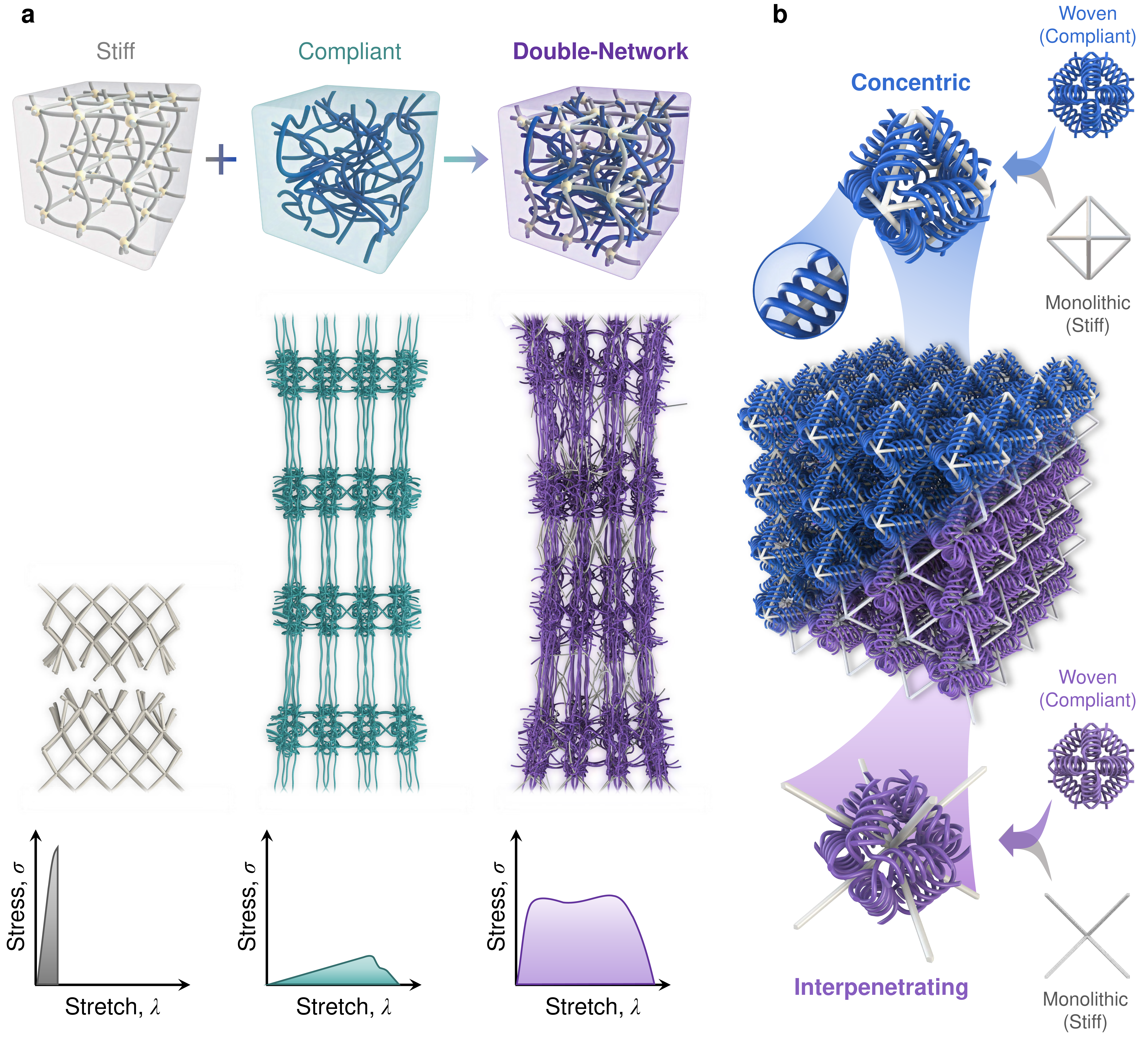}
    \caption{\textbf{Double-network-inspired (DNI) metamaterials.} \textbf{a}, Schematic of a conventional double network (DN) hydrogel and a stretched DNI metamaterial compared to their stiff (monolithic) and compliant (woven) counterparts. The corresponding stress-stretch curves illustrate the qualitative response of the DNI metamaterial compared to its counterparts. \textbf{b}, Illustration of the two design paradigms for DNI metamaterials presented in this study, namely concentric and interpenetrating designs.}
    \label{fig:design}
\end{figure}

\hfill\\
The design of DNI metamaterials revolves around the integration of monolithic-truss and woven architectures to mimic
the mechanical responses of both stiff and soft polymer networks. Specifically, we designed each network to be structurally independent and self-supporting, while becoming intertwined upon spatial superposition. 
To maximize the mechanical response tunability in the resulting DNI metamaterials, we developed two design paradigms that combine monolithic and woven geometries: (\emph{i}) the concentric design, wherein woven helices wrap around monolithic-lattice struts; and (\emph{ii}) the interpenetrating design, where woven and monolithic components form superposed reciprocal lattices (Fig.~\ref{fig:design}, S1, and Supplementary Information S1). 
Through leveraging of layer-by-layer additive manufacturing principles, both design paradigms allow for simultaneous fabrication of structurally independent stiff and compliant networks, bypassing the need for a multi-step, multi-material process such as the one conventionally employed in the fabrication of DN hydrogels.\cite{gong2003double, chen2015fundamentals}

\begin{figure}[H]
    \centering
    \includegraphics[width=0.85\textwidth]{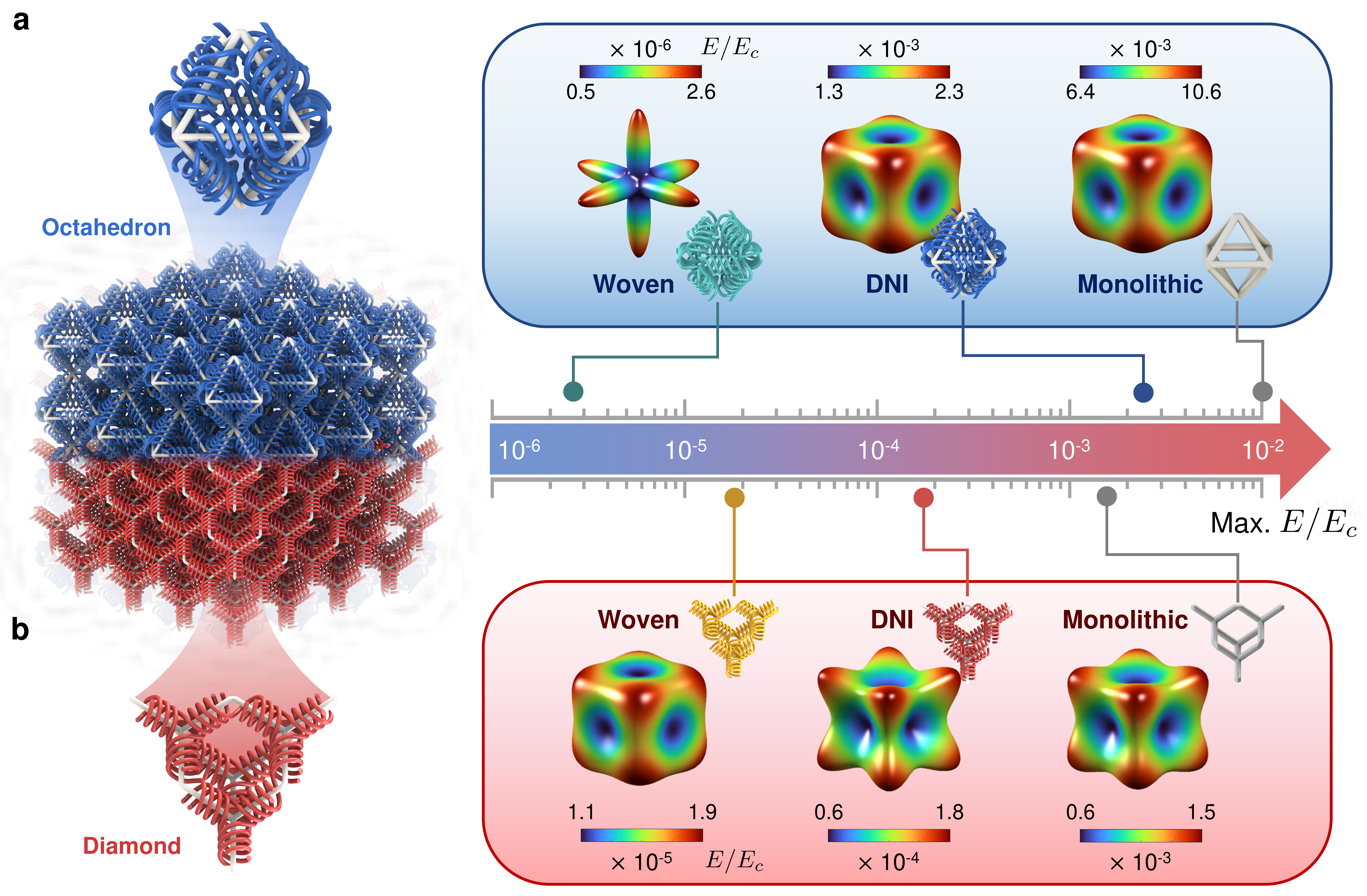}
    \caption{\textbf{Directional stiffness of DNI metamaterials.} Elastic surfaces of \textbf{a}, octahedron-based and \textbf{b}, diamond-based DNI architectures compared to their pure monolithic and woven counterparts at equivalent relative densities of 3.6$\pm$0.2\%, demonstrating the notable enhanced stiffness of the DNI architectures compared to a pure woven architecture. The Young's modulus of the architecture $E$ is normalized against that of the constituent material $E_{c}$.}
    \label{fig:elastic}
\end{figure}

To evaluate both DNI paradigms, we selected two  architectures---namely the octahedron and the diamond lattices---as representative morphologies of classical mechanical metamaterials. The octahedron was chosen to represent kinematically rigid architectures (stretching-dominated) while the diamond lattice was selected as a representative kinematically non-rigid architecture (bending-dominated). To determine the mechanical advantages of all DNI design combinations, we first characterized their linear-elastic responses via computational homogenization. Through enforcing periodic boundary conditions on a set of linearly independent imposed strains, we computed the elastic tensor of each architecture and represented the omnidirectional normalized stiffness as elastic surfaces, each of which directly conveyed an architecture's degree of anisotropy (Fig.~\ref{fig:elastic}).

In the case of the concentric octahedron architecture (interpenetrating-design results presented in Fig.~S3), a considerable change in anisotropy is observed between the stretching-dominated monolithic design and its woven counterpart, the latter of which is inevitably dominated by torsion and bending of helical segments \cite{moestopo2020pushing} (Fig.~\ref{fig:elastic}a). Therefore, the emergence of a monolithic network in a DNI octahedron design significantly alters the anisotropy to qualitatively resemble that of the monolithic architecture, owing to higher strain energy being localized in its stiff monolithic network than in the compliant woven network. 
In contrast, the concentric diamond architecture qualitatively maintains the anisotropic features of both its monolithic and woven networks, given that both are characterized by a bending-dominated deformation (Fig.~\ref{fig:elastic}b).

While the pure woven design in both octahedron and diamond architectures achieves stiffness two to three orders of magnitude  lower than their equal-mass monolithic counterparts in every direction, the resulting DNI architectures---comprising of approximately 20\% monolithic and 80\% woven components by volume---achieve stiffness within one order of magnitude of that achieved by the monolithic architectures. This decreased knock-down in stiffness suggests that only a small fraction of monolithic architecture suffices to achieve orders-of-magnitude enhancement in stiffness compared to classical woven architectures. These observations draw parallels to the behavior of a DN hydrogel, which can achieve specific stiffness levels akin to a stiff hydrogel despite being predominantly composed of a compliant network.\cite{gong2003double, sun2012highly}

\subsection{Dissipation through contact and entanglements}

\begin{figure}[H]
    \centering
    \includegraphics[width=\textwidth]{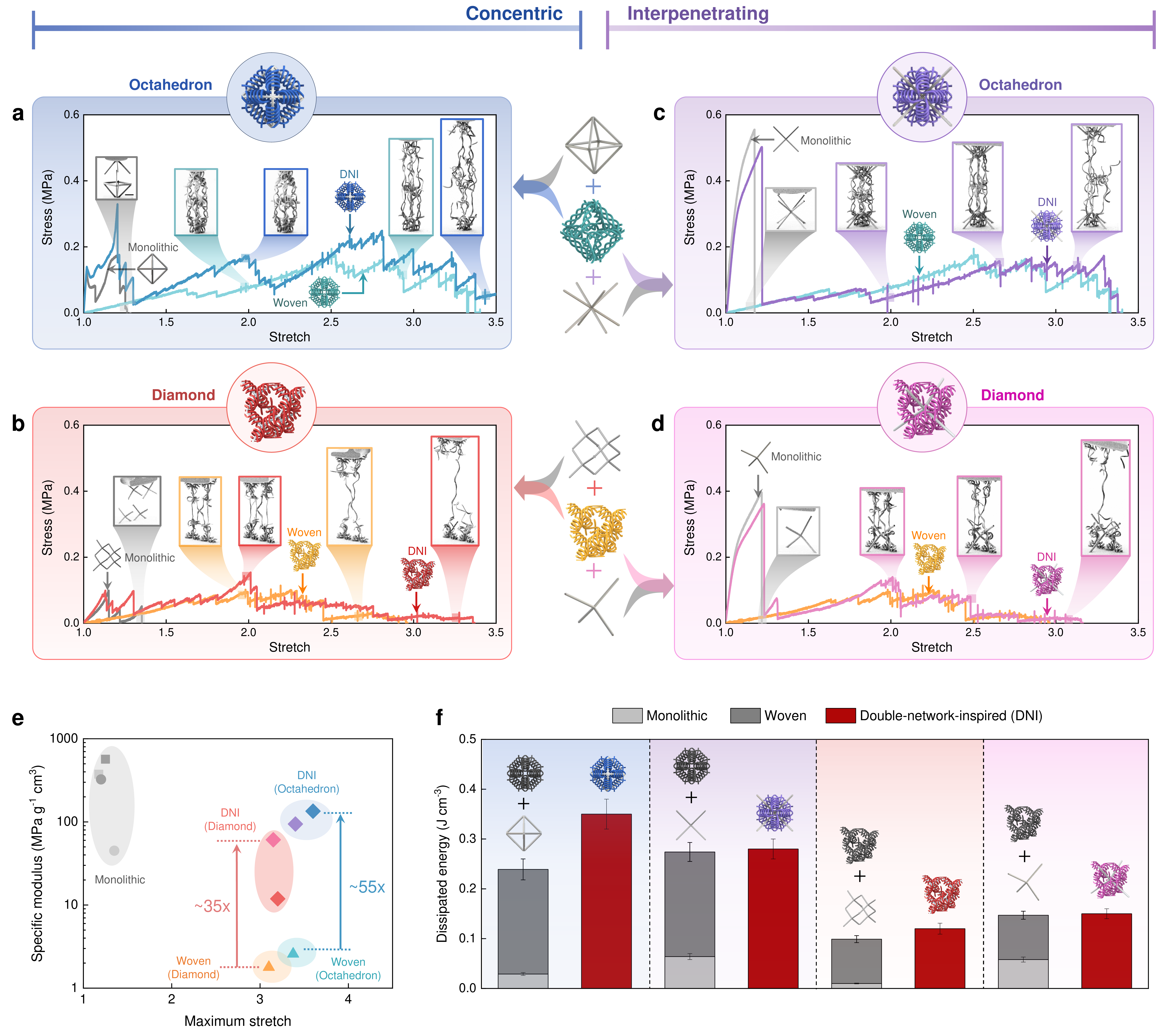}
    \caption{\textbf{\textit{In situ} uniaxial tension of DNI unit cells.} Stress-stretch curves of \textbf{a}, concentric octahedron; \textbf{b}, concentric diamond; \textbf{c}, interpenetrating octahedron; and \textbf{d}, interpenetrating diamond DNI unit cells along with their independent woven and monolithic components, tested under uniaxial tension. Scanning electron micrsocope (SEM) images show the deformed unit cells (UCs) at different stretch levels as indicated on the relevant curves. \textbf{e}, Comparison of specific modulus versus maximum achievable stretch for monolithic, woven, and DNI UCs, demonstrating the drastically increased modulus of the DNI UCs compared to pure woven UCs without compromising stretchability. \textbf{f}, Dissipated energy of DNI UCs compared to the sum of their constituent networks (monolithic and woven) when tested independently.}
    \label{fig:uc}
\end{figure}

\hfill\\
To determine the nonlinear mechanical properties and the resulting nonlinear design space of DNI metamaterials, we performed \emph{in situ} tension experiments on DNI unit cells (UCs) based on octahedron and diamond lattice geometries. To understand the contributions of each network separately, we additionally fabricated and tested the independent monolithic and woven networks, each with relative density of 0.6--1.1\% and 3.6--3.9\%, respectively. Superposition of the two networks led to the resulting DNI architectures with relative densities of 4.6\% and 4.5\% for the octahedron and diamond designs, respectively (Fig.~\ref{fig:uc}). We fabricated all UCs out of IP-Dip photoresist using a two-photon lithography process (Nanoscribe GmbH), and performed \emph{in situ} uniaxial tension experiments inside a scanning electron microscope using a silicon microgripper attached to a displacement-controlled nanoindenter (Alemnis AG, see Methods).

In all experiments, the explored DNI geometries achieved a stiffness comparable to their monolithic UCs while preserving the high stretchability of the woven UCs, attaining ultimate tensile stretches of up to 3.5 prior to a drop load to zero (Fig.~\ref{fig:uc}a--d). Conversely, pure monolithic UCs exhibited complete failure at considerably lower stretch values between 1.2 to 1.4. For pure woven UCs, while they sustain deformation to stretch values exceeding 3, their specific Young’s modulus remained more than an order-of-magnitude lower than that of monolithic and DNI UCs (Fig.~\ref{fig:uc}e). Furthermore, the dissipated energy density of the concentric DNI UCs---defined as the area under the stress-stretch curve---surpassed the sum of the dissipated energy densities of their monolithic and woven components. The difference is particularly evident in the octahedron geometry, which constituted a 46\% increase in the concentric case (Fig.~\ref{fig:uc}f, S5, and Supplementary Information S3). 
In contrast, the energy dissipation of the interpenetrating DNI UCs is similar to the sum of their counterparts. A closer examination of the deformation behavior of the concentric DNI UCs revealed multiple contact points between the monolithic and woven components, whereas little to no self-contact was observed in the interpenetrating DNI UCs. These observations suggest that the heightened energy dissipation in the DNI UCs can be attributed to increased frictional dissipation arising from contact interactions such as entanglements between the monolithic and woven components.

\begin{figure}[H]
    \centering
    \includegraphics[width=0.9\textwidth]{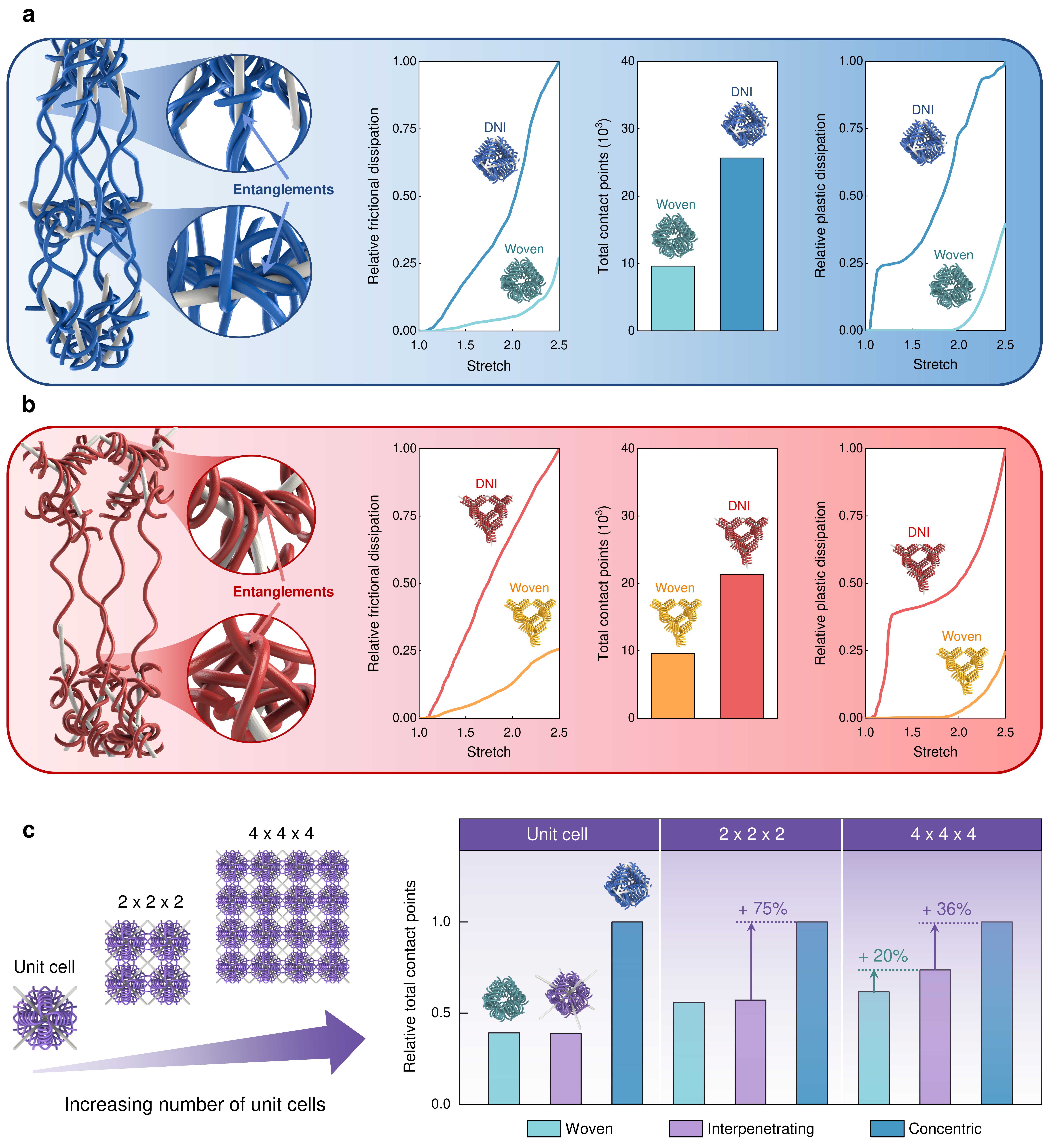}
    \caption{\textbf{Nonlinear finite element modeling of DNI architectures.} Deformed configuration (extracted from finite element models) of DNI UCs showing representative locations of entanglements as well as energy dissipation metrics such as relative frictional dissipation, total number of contact points accumulated up to a stretch of 2.5, and relative plastic dissipation for \textbf{a}, concentric octahedron and \textbf{b}, concentric diamond DNI architectures. \textbf{c}, Progression in
    relative contact points accumulated up to stretch of 2.5 for pure woven, interpenetrating octahedron DNI, and concentric octahedron DNI upon increasing the lattice tessellation size, revealing the increased difference in total contact points between interpenetrating DNI and pure woven but decreased difference between concentric DNI and interpenetrating/pure woven architectures upon increasing the number of UCs tessellated.}
    \label{fig:fem} 
\end{figure}

To quantitatively determine the role of contact and entanglements on the energy dissipation capability of DNI architectures, we implemented nonlinear finite element models that accounted for frictional contact and large deformation (Fig.~\ref{fig:fem}). To ensure accurate geometric representation of the geometries, these morphologies were meshed with quadratic tetrahedral elements, ensuring a minimum of 20 elements spanning the diameter of a beam or fiber. To account for material nonlinearity, we implemented a strain-hardening material model based on the experimental stress-strain curves of the constituent polymer (Fig.~S2), while the failure of the monolithic components was modeled via cohesive surface interactions at the lattice nodes (Methods). The validity of these computational models was confirmed by a high degree of congruence in the kinematics and failure of the architectures, when compared to the \emph{in situ} experimental observations (Fig.~S6) under the same tensile boundary conditions. A detailed comparison of the energy dissipation mechanisms between concentric DNI and purely woven UCs (Fig.~\ref{fig:fem}a and b), reveals that the DNI UCs significantly outperform purely woven UCs in terms of frictional energy dissipation. This superior performance is attributed to the enhanced contact area and the number of contact points (Fig.~S7), a direct consequence of the intricate entanglements formed between the woven fibers and the monolithic struts. 
The number of contact points is defined by the number of paired elements with nonzero contact pressure, while the total contact points represent the accumulated number of contact points from each stretch increment and up to a stretch of 2.5.
Furthermore, the disparity in the total number of contact points is more pronounced between the octahedron-based DNI UCs and woven UCs than between their diamond-based counterparts. This distinction offers a partial explanation for the heightened enhancement in energy dissipation observed in octahedron-based DNI UCs relative to the diamond-based DNI UCs (Fig.~\ref{fig:uc}f), underlining the critical role of network topology in optimizing energy dissipation. Besides frictional dissipation via contact sliding and entanglements, the earlier onset of plastic dissipation induced by buckling of the monolithic network in the DNI UCs gives rise to enhanced strength and energy dissipation compared to pure woven UCs at lower stretch values. These observations highlight the prominent role of the monolithic network in enhancing the overall performance of DNI architectures.

While the DNI architectures exhibit superior mechanical performance over purely woven architectures at the unit cell level, we sought to determine if these advantageous properties persist across larger tessellations within a metamaterial framework. However, extending nonlinear FEM analyses to accommodate large stretch values using 3D tetrahedral solid elements for these complex woven and DNI geometries proves to be computationally demanding.
To address the computational-cost challenge of modeling large tessellations using the tetrahedral-element representation described above, we implemented alternative beam-element representations of the architectures while still capturing large-deformation and nonlinear responses (see Methods). This structural-element representation of the architectures agreed kinematically and quantitatively against the continuum-element representations in terms of stress-stretch responses, curvature patterns, and density of contact points, while enabling computation of arbitrarily large DNI tessellations (Fig.~S8).  
Following validation of the models, we evaluated the extent of the network's entanglements through an analysis of the cumulative number of contact points, specifically by assessing the total number of contact points as a function of the tessellation size for both woven and DNI (concentric and interpenetrating) architectures (Fig.~\ref{fig:fem}c). Notably, as the number of unit cells increases, the difference in the total number of contact points between a purely woven and an interpenetrating DNI architecture widens (from 2\% to 20\%), whereas the gap between the concentric and interpenetrating DNI configurations narrows (from 75\% to 36\%). This trend suggests a marked difference in the deformed morphologies and failure mechanisms of the DNI architectures at larger tessellations.

\subsection{Failure delocalization via engineered defects}
\begin{figure}[H]
    \centering
    \includegraphics[width=0.65\textwidth]{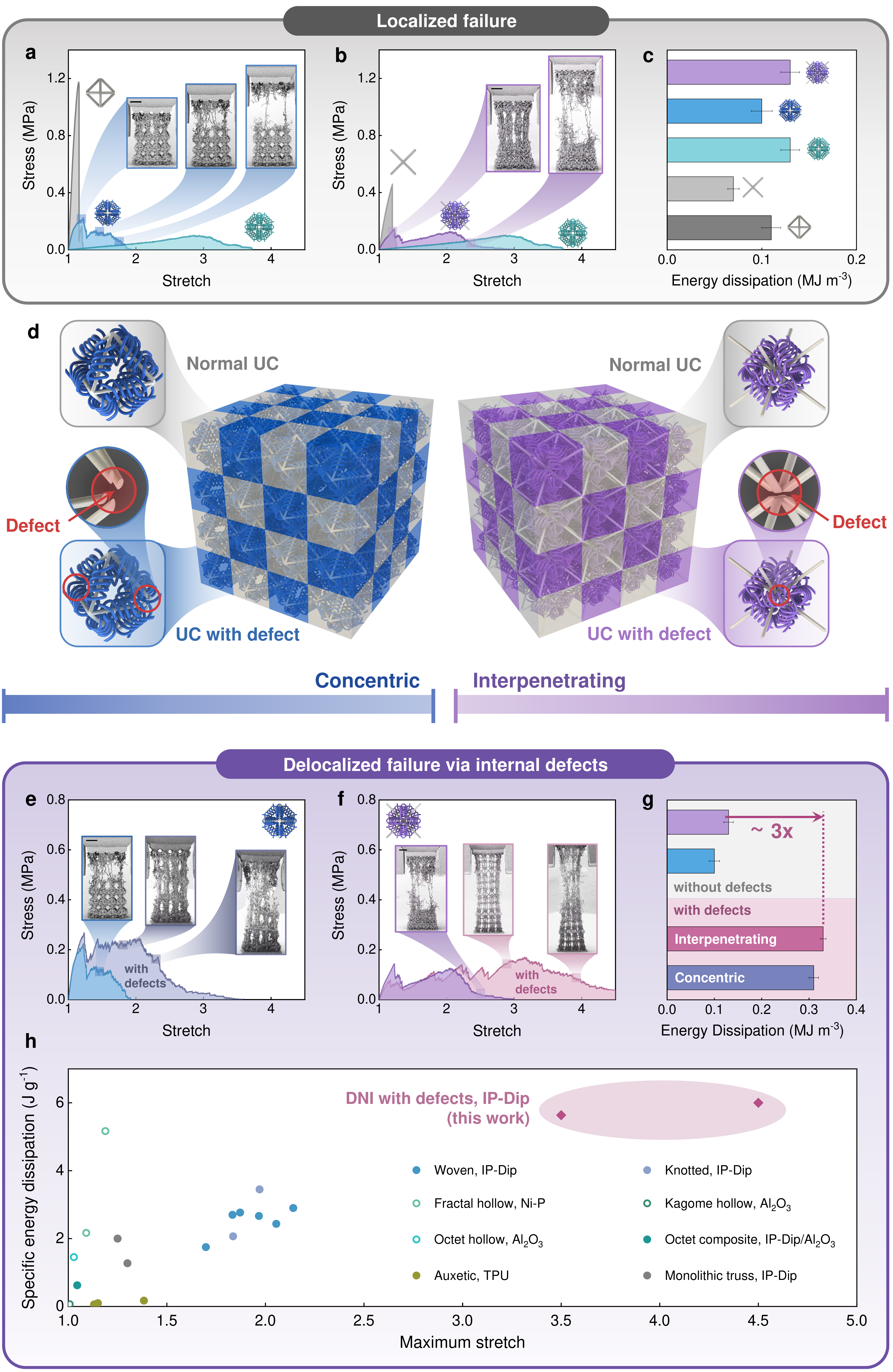}
    \caption{\textbf{Delocalizing failure in DNI metamaterials via engineered internal defects.} Stress-stretch curves of monolithic and woven metamaterials compared to \textbf{a}, concentric and \textbf{b}, interpenetrating DNI metamaterials under uniaxial tension at a relative density of 4.6\%, demonstrating the localized failure and reduced maximum stretch of the DNI metamaterials compared to the pure woven metamaterials. \textbf{c}, Corresponding energy dissipation per unit volume by the monolithic, woven, and DNI metamaterials. \textbf{d}, Schematic of engineered internal defect incorporation into concentric and interpenetrating DNI metamaterials following a checkerboard-like pattern. Stress-stretch curves of \textbf{e}, concentric and \textbf{f}, interpenetrating DNI metamaterials with engineered internal defects compared to those without, demonstrating drastically enhanced stretchability. \textbf{g}, Comparison of energy dissipation of DNI metamaterials with and without defects. \textbf{h}, Performance comparison of DNI metamaterials with defects against reported metamaterials under tension.\cite{moestopo2020pushing, moestopo2023knots, zheng2016multiscale, montemayor2016insensitivity, mateos2019discrete, bauer2015push, hu2024compressive} Scale bars, 50 \textmu{}m.}
    \label{fig:defect} 
\end{figure}

\hfill\\
While the advantages of DNI architectures were demonstrated experimentally and numerically on individual unit cells, we set out to determine the translation of these benefits to finite tessellations of DNI metamaterials. We performed \emph{in situ} uniaxial tension experiments on monolithic, pure woven, and DNI architectures, each with a $4 \times 4 \times 4$ tessellation at equivalent relative densities of $4.6\pm 0.3\%$, resulting in sample sizes of 240 $\times$ 240 $\times$ 240 \textmu{}m$^3$. These experiments revealed a pronounced localized failure in the DNI metamaterials, driven by failure of the monolithic network, which was primarily concentrated within a single unit cell layer of the metamaterial. This specific localization led to concentrated deformation within the woven network as well, with only woven unit cells in that layer undergoing large stretches. This phenomenon is attributed to the interactions between the intact portions of the monolithic network and the woven network outside the fracture zone, which result in the deformation being confined to a specific region, effectively isolating the failure and reducing the overall stretchability of the DNI metamaterial tesssellations (Fig.~\ref{fig:defect}a, b, and S9). The occurrence of localized failure within DNI metamaterials led to a reduction dissipated energy density (Fig.~\ref{fig:defect}c), resulting in the pure woven metamaterials demonstrating similar performance to interpenetrating DNI metamaterials, both of which outperformed the concentric DNI metamaterials by 30\%. These nonlinear responses contrast with the behavior observed in individual unit cells, implying a reduced effectiveness of the DNI design strategy when tessellating the same UC across larger volumes.

To distribute damage more evenly across the DNI metamaterials, we engineered internal defects into the monolithic network of the DNI UCs. The metamaterial was then assembled from alternating UCs, some with defects and some without, creating a 3D checkerboard-like pattern (Fig.~\ref{fig:defect}d). We conducted \emph{in situ} uniaxial tension experiments on DNI metamaterials with engineered internal defects, which demonstrated a significant improvement in stretchability compared to those lacking such defects. In contrast to DNI metamaterials comprising identical repeating unit cells, which tend to undergo localized deformation upon monolithic network failure, those with internal defects demonstrate a distinctive behavior characterized by homogeneous deformation throughout (Fig.~\ref{fig:defect}e and f). In this scenario, all UCs experience uniform deformation up to the point of failure in the woven network. Notably, the maximum stretch significantly increased from approximately 2 to 3.5 for the concentric DNI metamaterial and from 3 to 4.5 for the interpenetrating design. Nonlinear finite element models of the DNI metamaterials with internal defects also reveal a notable absence of localized stress concentration, unlike their defect-free counterparts (Fig.~S10). 
This phenomenon stems from the occurrence of monolithic failure distributed across every layer within the metamaterial, inducing more uniformly distributed regions characterized by entanglements between the fractured monolithic and stretched woven networks which dissipate additional energy (Fig.~S10). Consequently, DNI metamaterials with internal defects exhibit a three-fold increase in energy dissipation compared to their defect-free counterparts (Fig.~\ref{fig:defect}g), representing a substantial improvement over pure monolithic and woven materials and surpassing the performance observed in individual UCs. These findings underscore the superior effectiveness of the defect-incorporated DNI design approach, particularly when applied to larger metamaterial tessellations.

We compared the tensile performance of our DNI metamaterials with engineered internal defects against previously reported metamaterials in terms of maximum stretch and specific energy dissipation---defined as energy density normalized by the mass density of the material (Fig.~\ref{fig:defect}h). Most metamaterials under tension, except for woven and knotted architectures, fail at stretch values below 2 due to stress concentration at their nodes, regardless of their constituent materials, limiting their energy dissipation performance. Although woven and knotted architectures can achieve stretch values greater than 2, their low stiffness restricts energy dissipation at lower stretch values before fiber-fiber entanglements and stretching occur. In contrast, DNI metamaterials exhibit superior specific energy dissipation and stretchability due to the monolithic network providing initial high stiffness and strength at low stretch values (below 1.5) followed by entanglements between fractured monolithic fragments and the woven network. These entanglements aid in maintaining high-stress levels as the woven fibers are uncoiled and stretched. Importantly, the incorporation of engineered internal defects in the monolithic network ensures failure distribution across every layer within the metamaterial, preventing localization to the initial fracture region.

\subsection{Defect tolerance and fracture energy}
\begin{figure}[H]
    \centering
    \includegraphics[width=1\textwidth]{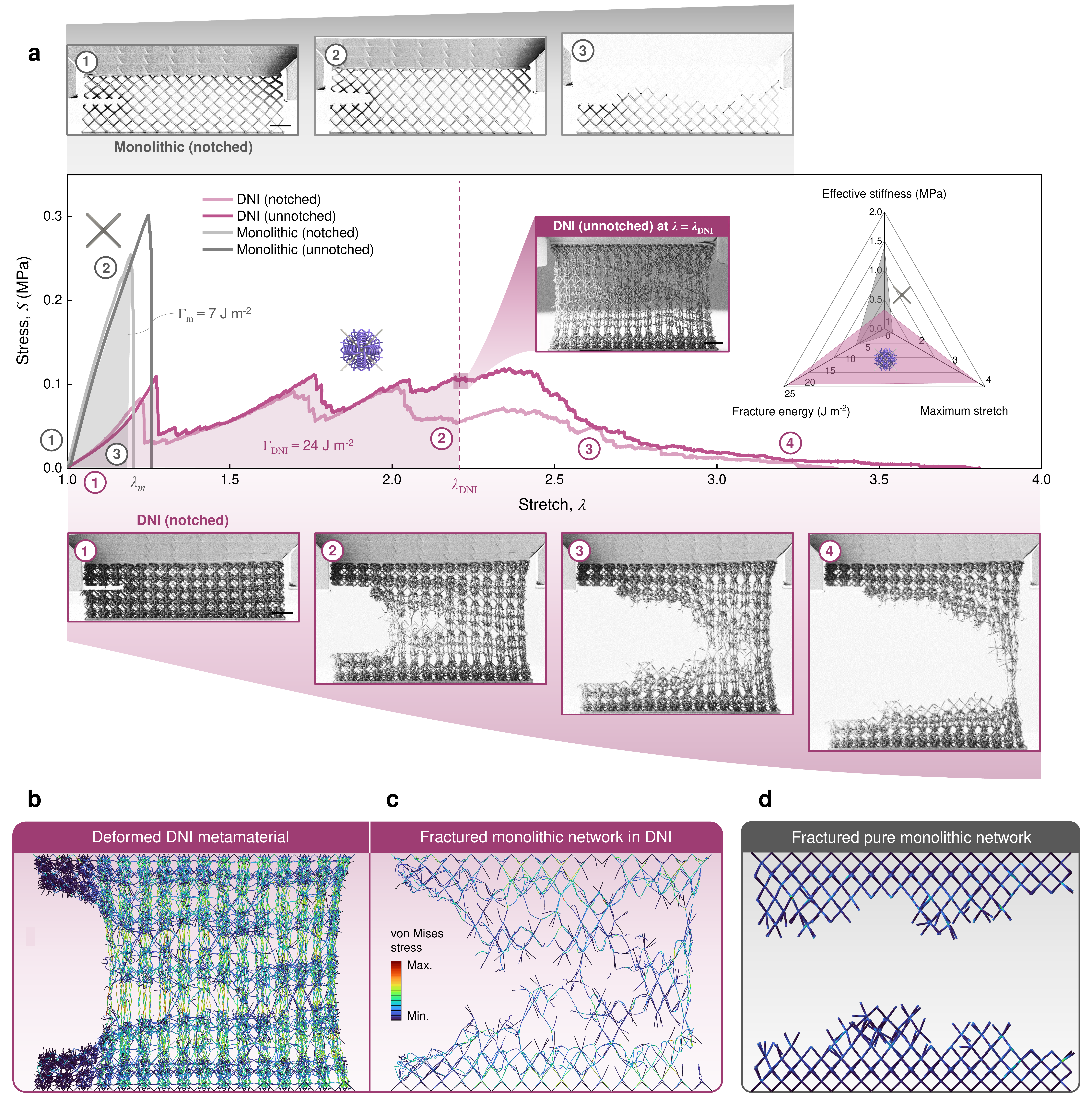}
    \caption{\textbf{Pure-shear experiments on monolithic and DNI metamaterials.} \textbf{a}, Nominal stress-stretch curves of  monolithic and DNI metamaterials (notched and unnotched) along with SEM micrographs taken during the pure-shear experiments. \textbf{b}, Deformed morphology of the notched DNI metamaterial obtained via nonlinear beam-element simulation; and \textbf{c}, the corresponding isolated fractured monolithic network in the deformed DNI morphology, showing a large damage zone characterized by failure at multiple locations. \textbf{d}, Simulated fractured network in the notched pure monolithic metamaterial, showing failure only in the vicinity of the crack tip. Scale bars, 100 \textmu{}m.}
    \label{fig:fracture} 
\end{figure}

\hfill\\
To probe the fracture response of the DNI metamaterials in contrast to conventional monolithic strut-based metamaterials, we conducted pure-shear-like fracture experiments inspired by the Rivlin-Thomas model (Fig.~\ref{fig:fracture}),\cite{rivlin1953rupture} which is commonly used to determine the fracture energy of tough hydrogels and hyperelastic materials\cite{sun2012highly, zhao2014multi, zhang2015predicting, zheng2021chain, liu20243d} (Supplementary Information S4 and Methods). Briefly, the procedure involves (i) stretching a notched sample until the crack starts to propagate, and noting the critical stretch of the sample $\lambda_c$ (represented by $\lambda_{\text{DNI}}$ and $\lambda_{\text{m}}$ for the DNI and monolithic samples, respectively); and (ii) stretching an unnotched sample and integrating its nominal stress $S$ from $\lambda=1$ to $\lambda_c$, and multiplying it by the original height of the sample $H_0$, resulting in the fracture energy
\[\Gamma = H_0 \int_{1}^{\lambda_c} S \text{d}\lambda . \]
Snapshots captured during the \emph{in situ} fracture experiments unveiled distinct behaviors in the tested samples. The monolithic specimens exhibited localized and abrupt failure patterns as the crack propagated, contrasting with the notched DNI sample, which displayed notable crack blunting before eventual crack propagation across the specimen (Fig.~\ref{fig:fracture}a). This divergence led to a substantial enhancement in the critical stretch, from approximately 1.2 in the monolithic sample to 2.2 in the DNI variant. Consequently, the fracture energy of the DNI metamaterial surpassed that of the monolithic counterpart by a factor of three at comparable relative densities. 

To gain deeper mechanistic understanding of this enhanced failure response, we modeled these fracture experiments using the previously described nonlinear beam-element computational representations of the metamaterials (Fig.~\ref{fig:fracture}b--d, S11, and S12).
In the notched DNI metamaterial simulation, stress concentration around the crack tip during stretching led to eventual rupture of the stiff, brittle monolithic network at that juncture. As the monolithic struts failed, the load was distributed to the woven network, causing the fibers to extend. Subsequent strain hardening within the woven network---stemming from fiber uncoiling, entanglements with broken monolithic fragments, and inter-fiber interactions---led to further rupturing of neighboring monolithic struts, expanding the damage zone and blunting the crack. This blunting response mitigated stress concentration at the crack tip, resulting in impeded crack propagation. Further stretching of the sample continued this iterative process, progressively enlarging the damage zone until the tensile stress at the crack tip surpassed a critical threshold at a stretch of $\lambda_{\text{DNI}}$. Beyond this point, the woven fibers progressively failed, initiating crack propagation akin to typical DN hydrogels.\cite{zheng2021chain} 

The crack dynamics of DNI metamaterials under pure-shear also draw parallels to DN hydrogels, whereby a distinctive ``stick-slip" phenomenon manifests.\cite{zheng2024unique} The slip phenomenon can be linked to the quick, dynamic fracture occurring within the damage zone, whereas the stick phenomenon, which delays crack propagation, can be associated with developing a new damage zone at the crack tip. Establishing this new damage zone at the crack tip thus requires cumulative internal fracturing of the rigid, monolithic network, thereby demanding additional strain energy input to propagate the crack tip. In contrast, the relatively low fracture energy of the monolithic single-network metamaterial parallels that of hydrogels with covalent crosslinks.\cite{gong2003double, sun2012highly, kim2021fracture, li2023crack} Specifically, stretching the notched monolithic metamaterial yields non-uniform deformation, with the region directly ahead of the notch experiencing greater strain compared to the rest of the sample. Thus, for crack propagation, only localized failure of struts at the notch is required.\cite{sun2012highly, kim2021fracture, li2023crack} Owing to this delocalization of failure, DNI metamaterials exhibit superior defect tolerance compared to traditional truss-based counterparts, yielding substantially enhanced failure strains and fracture energies.

\subsection{Summary and outlook}
\hfill\\
In summary, we introduced a general design framework for mechanical metamaterials that draw inspiration from DN hydrogels, ultimately achieving both high stiffness and stretchability (exceeding 4). Unlike previous approaches that focus solely on multistability or fracture for energy dissipation, our method harnesses and optimizes frictional dissipation via entanglements in DNI metamaterials, resulting in ultrahigh energy dissipation surpassing the combined capabilities of their woven and monolithic counterparts. Counterintuitively,  we demonstrated that engineered internal defects within the DNI metamaterials led to a significant increase in energy dissipation, contrary to conventional wisdom where defects often degrade mechanical properties. This enhancement yields over a threefold increase in energy dissipation compared to pure woven and monolithic metamaterials, resulting in a unique combination of ultrahigh specific energy dissipation and stretchability. Fracture experiments and nonlinear computational models with frictional contact and failure further demonstrate defect tolerance of DNI metamaterials compared to conventional monolithic truss metamaterials, while exhibiting analogous responses to DN hydrogels.

This design framework opens avenues for developing polymer-inspired metamaterials with unique mechanical responses applicable across various engineering domains. For example, in the creation of conformal bio-scaffolds and implants mimicking the diverse characteristics of living tissues,\cite{butcher2009tense} enhancing scaffold stiffness without compromising deformability is crucial.\cite{wang2022bioinspired} Similarly, in soft pressure sensor applications, materials with adjustable stiffness but retained stretchability can mitigate the sensitivity-pressure range trade-off inherent in existing sensors.\cite{luo2023technology} Moreover, given the analogous mechanical responses between DNI metamaterials and DN hydrogels (e.g., energy dissipation mechanisms, crack dynamics), this framework can serve as a valuable tool for understanding the complex mechanics of existing and future polymer networks,\cite{zhao2014multi, gu2019macro} especially for polymers where entanglements play a central role.\cite{kim2021fracture} Hence, we envision this framework providing researchers with a foundation to efficiently optimize the topology of complex polymer networks and inspire the development of new polymer systems.

\subsection{Methods}
\hfill\\
\emph{Fabrication of DNI metamaterials}
\hfill\\
All specimens were fabricated on silicon (Si) substrates using IP-Dip photoresist, an acrylate-based polymer, through two-photon polymerization using a Photonic Professional GT2 system (Nanoscribe GmbH). Laser power and scan speed settings of 32.5 mW and 10 mm s$^{-1}$ were applied for the architected components, while the monolithic parts (i.e., the tensile fixtures atop the architecture) were printed at a laser power of 25 mW and a scan speed of 10 mm s$^{-1}$ to minimize cavitation. These parameters were held constant during printing, ensuring uniformity in the fabrication process. This included maintaining a consistent hatching and slicing distance of 0.2 µm for all architected components and pillars, while the tensile fixtures were printed with a hatching and slicing distance of 0.2 µm and 0.3 µm, respectively. Following printing, specimens underwent immersion in propylene glycol monomethyl ether acetate for approximately 30 minutes for unit cells/pillars and at least 4 hours for larger tessellations to prevent excess resin entrapment. Subsequently, a 5-minute rinse in isopropanol was conducted, followed by drying using a critical point dryer (Autosamdri 931, Tousimis). The samples were then coated with a 10 nm gold (Au) layer via sputtering (SCD 040, Balzers) for appropriate visualization during \emph{in situ} mechanical testing. The relative density of each lattice was assessed by analyzing the geometric characteristics of individual unit cells using SEM (Gemini 450, ZEISS), which involved measuring parameters such as unit-cell size and strut dimensions. Subsequently, the unit cell was digitally recreated in a CAD software (SolidWorks, Dassault Systèmes) to determine its volume. This volume was then compared to the bounding-box volume of the unit cell to calculate the relative density. Focused ion beam milling (FIB, FEI Helios Nanolab 600) was employed to cut the notch on samples for pure-shear experiments.

\noindent\emph{In situ mechanical characterization}
\hfill\\
Uniaxial tension experiments were conducted using a custom-designed tension gripper affixed to a nanoindenter (Alemnis AG) placed inside the SEM (Gemini 450, ZEISS) to facilitate real-time imaging of the experimental events. The gripper was operated in a displacement-controlled mode, and all specimens were loaded at a prescribed strain rate of approximately $5 \times 10^{-2} \, \text{s}^{-1}$. The stress-strain data was derived by normalizing the load-displacement data with the nominal area and height of each specimen, respectively. Compliance correction was performed to exclude the compliance of the system other than the sample (see Supplementary Information S2 for details). The dissipated energy density for each sample was obtained by integrating the area under the stress-stretch curve. 

\noindent\emph{Finite element modeling}
\hfill\\
All finite element simulations were conducted using ABAQUS (Simulia). Computational homogenization was employed to determine the directional stiffness of the monolithic, woven, and DNI architectures using ABAQUS/Standard (Simulia). Each architecture was meshed with quadratic tetrahedral (C3D10) elements utilizing full integration. The mechanical properties of the IP-Dip samples were obtained from uniaxial tension experiments conducted on printed monolithic IP-Dip pillars (see Supplementary Information S2 for details). Periodic boundary conditions were enforced, while imposing an average strain on the unit cell.
A linear perturbation was applied sequentially to the unit cells along six linearly independent strain vectors. The stress corresponding to each applied strain case was determined, and the compliance tensor was subsequently constructed.

To study the kinematic behaviors and energy dissipation mechanisms inherent in both DNI and woven architectures, dynamic explicit simulations were conducted using ABAQUS/Explicit. The simulations employed either tetrahedral (C3D10) or linear beam (B31) elements depending on the problem size. Due to computational constraints, tetrahedral elements were utilized solely for modeling individual unit cells, while beam elements were employed for modeling both unit cells and larger tessellations. General contact modeling was adopted to simulate interactions between woven-woven fibers and between woven fibers and monolithic struts. Cohesive surface interactions were implemented at locations of high curvature/stress concentration (i.e., lattice nodes) to approximate the failure of the monolithic component in tetrahedral element simulations, whereas strain-based element deletion was utilized for beam-element simulations to capture the kinematic behaviors observed in the experimental results (see Fig.~S6, S8, and S10 for more details).

\subsection{Data availability}
\hfill\\
The data that support the findings of this study are present in the paper and/or in the Supplementary Information. Additional data related to the paper are available from the corresponding author upon request.

\subsection{Acknowledgements}
\hfill\\
This work was performed in part in the MIT.nano Fabrication and Characterization Facilities. The authors thank S. Dhulipala for his assistance with linear finite element simulations and tension experiments.

\subsection{Author contributions}
\hfill\\
C.M.P. and J.U.S. conceived this study. J.U.S. designed the architectures, fabricated samples, performed mechanical experiments, analyzed data, conducted part of the simulations. B.F.G.A. and M.C. performed simulations and analyzed data. C.M.P. supervised the project. J.U.S and C.M.P. wrote the manuscript with input from all authors.

\subsection{Competing interests}
\hfill\\
The authors declare no competing interests.

\nolinenumbers
\vspace{10pt}

\subsection{References}
{\spacing{1}
\bibliography{references}
}
\bibliographystyle{naturemag.bst}

\end{document}


\maketitle

\begin{affiliations}
\footnotesize
 \item Department of Mechanical Engineering, Massachusetts Institute of Technology, Cambridge, MA 02139, USA
\end{affiliations}
\vspace{10pt}
\spacing{1.0}


\subsection{S1: Design space of double-network-inspired (DNI) metamaterials}
\hfill\\
 In this work, we explored two design paradigms of DNI metamaterials, namely the (i) concentric, and (ii) interpenetrating designs (Fig.~\ref{fig:designspace}a). These DNI designs are based on two types of woven unit cells (UCs), octahedron-based and diamond-based. Unlike typical strut-based lattices which form monolithic junctions or nodes at regions where neighboring struts intersect, the woven unit cells are instead interwoven at the effective strut junctions. Each woven strut, with an effective radius $R$ and length $L$ consist of three helical fibers with fiber radius $r$ and pitch $\lambda$, that are extended beyond the struts to also make up the three closest neighboring woven struts. For this work, the pitch of the woven fibers $\lambda$ is fixed with respect to the length of each woven strut following $L = 4/3\lambda$. Therefore, the relative density of a woven UC $\bar{\rho}_w$, depends on three primary variables $r$, $R$, and $\lambda$. To create the DNI unit cells, each of the woven unit cells are combined with two types of monolithic strut-based unit cells with strut radius $r^*$ and strut length $L^*$, to form the four different types of DNI geometries investigated in this work (Fig.~\ref{fig:designspace}b). For the \emph{in situ} tension experiments on the DNI unit cells, the relative density of the DNI unit cells is given by $\bar{\rho}_{\text{DNI}}=\bar{\rho}_w+\bar{\rho}_s$, where $\bar{\rho}_s$ is the relative density of the monolithic UC. This superposition of relative densities demonstrates that the energy dissipated by the DNI unit cells is greater than the total energy dissipated by its woven and monolithic constituents independently. Subsequently, for the tension experiments involving $4 \times 4 \times 4$ tessellations of the UCs, as well as the pure-shear experiments, the relative density of the DNI, woven, and monolithic metamaterials are designed to be equal; $\bar{\rho}_{\text{DNI}}=\bar{\rho}_w=\bar{\rho}_s$.

\subsection{S2: Compliance correction}
\hfill\\
For the \emph{in situ} tension experiments, fabricating a customized tensile fixture is required to allow the microgripper (which is attached to the nanoindenter) to hold onto the architected lattice sample. A base plate was also fabricated underneath each architecture to enhance adhesion to the Si substrate and prevent delamination of the samples upon tension. This tension setup is analogous to applying tension load on a series of springs (Fig.~\ref{fig:compliance}a). However, as the tensile fixture and base plate are made of IP-Dip, which is the same constituent material as the lattice, the compliance of the tensile fixture $C_{\text {fixture }}$ and base plate $C_{\text {base }}$ cannot be assumed to be infinitely small compared to the compliance of the architected sample $C_{\text {architecture }}$.\cite{moestopo2020pushing} Therefore, the displacements of the tensile fixture $u_{\text {fixture }}$ and base plate $u_{\text {base }}$ must be deducted from the total displacement $u_{\text {total }}$, giving the expression for the actual displacement of the architected sample $u_{\text {architecture }}$ as
\begin{equation}
u_{\text {architecture }}=u_{\text {total }}-u_{\text {fixture }}-u_{\text {base },}
\end{equation}
\begin{equation}
u_{\text {total }}=\left(C_{\text {architecture }}+C_{\text {fixture }}+C_{\text {base }}\right) \times \text { Load }.
\end{equation}
Note that, in this work, the compliance of the nanoindenter and Si substrate are assumed to be rigid with respect to the IP-Dip components owing to the large difference in stiffness (over 2 orders of magnitude). Using the stiffness-compliance relation, we can approximate the combined stiffness of both the tensile fixture and base as
\begin{equation}
K_{\text {fixture+base }}=\frac{1}{C_{\text {fixture }}+C_{\text {base }}}.
\end{equation}
The combined stiffness of the tensile fixture and base was obtained by performing uniaxial tension on the printed IP-Dip tensile fixture and base without the architected sample in between (Fig.~\ref{fig:compliance}a) and calculating the slope of the corresponding load-displacement curve. Finally, the actual displacement of the architected sample can be calculated as
\begin{equation}
u_{\text {architecture }}=u_{\text {total }}-\left(\frac{1}{K_{\text {fixture+base }}} \times \text { Load }\right).
\end{equation}
A comparison between the corrected and un-corrected load-displacement curve for a DNI UC (Fig.~\ref{fig:compliance}b) reveals negligible difference between the compliance of the architecture with the total compliance of the system, which is due to the low relative density of the architectures investigated (below 10\%).

\subsection{S3: Parametric study on DNI unit cells}
\hfill\\
To optimize the volume fraction of monolithic to woven network for maximal specific stiffness and energy absorption, we conducted a comprehensive parametric analysis on DNI unit cells. This involved varying the radius of the woven fiber \( r \) while maintaining a consistent monolithic strut radius \( r^* \) set at \( \SI{1.25}{\micro\meter} \) and effective radius of the woven strut \( R \) at \( \SI{6}{\micro\meter} \). Through \emph{in situ} uniaxial tension experiments, extended to complete failure across all monolithic, woven, and DNI UC variations, we calculated their specific modulus and energy dissipation (Fig.~\ref{fig:optimization}). Our findings indicate a consistent trend: lower woven fiber radius generally corresponds to higher specific modulus and energy dissipation for all DNI designs. This outcome can be attributed to the dominance of the monolithic network in achieving high stiffness. Consequently, increasing the relative density of the woven component decreases the volume fraction of the monolithic component within the total DNI UC volume, thereby reducing its specific stiffness. Furthermore, the diminishing specific energy dissipation of DNI unit cells with increasing woven fiber radius can be ascribed to the reduced stretchability of woven networks as the fiber radius enlarges. This is due to the earlier onset of contact and entanglements between the woven fibers, resulting in earlier yielding within the fibers and thus reduced fracture strain. However, reducing the woven fiber radius further poses manufacturing challenges due to the increased compliance of the woven network, leading to heightened defects between stitched regions. Moreover, based on the deformation behavior observed in DNI unit cells, such as the concentric octahedron (Fig.~\ref{fig:simvsexp}), a further reduction in the radius of the woven fiber may decrease the stretchability due to entanglements between the monolithic struts and the woven fibers. In particular, at a radius of woven fibers of \( \SI{1}{\micro\meter} \), entanglements between multiple woven fibers and a monolithic strut resulted in a secondary failure event of the monolithic network (see transition between frame 7 and frame 8 in Fig.~\ref{fig:simvsexp}). However, pushing the radius of the woven fiber even lower would weaken the woven fibers to the extent that entanglements between monolithic struts and woven fibers would lead to the failure of the woven fibers instead, compromising stretchability and energy dissipation. Thus, in the tensile testing of \( 4 \times 4 \times 4 \) tessellations of DNI metamaterials and pure-shear experiments, we opted for a woven fiber radius of \( \SI{1}{\micro\meter} \).

\subsection{S4: Selection of sample dimensions for fracture experiments}
\hfill\\
In the classical pure-shear experiments, samples are typically clamped along their long edges (i.e., along the width direction), with the width \( W \) significantly larger than the height \( H \) (i.e., the separation between the clamps) and the height considerably larger than the thickness \( T \) of the sample thus satisfying the relation
\begin{equation}
W \gg H \gg T.
\end{equation}
Under the assumption that the sample resembles an infinitely long strip with a semi-infinitely long edge crack (for fracture experiments) under quasi-static loading, increasing the crack length by an amount \text{d}$c$ does not alter the state of strain in the region near the crack tip but shifts this region parallel to the crack direction.\cite{rivlin1953rupture} Consequently, the region behind the crack tip grows at the expense of the region experiencing pure shear.\cite{rivlin1953rupture} Thus, an increase in the cut length \text{d}$c$ transfers a volume \( H \times T \times \text{d}c \) of the sample from a state of pure shear to the undeformed state.\cite{rivlin1953rupture} Theoretically, the value of intrinsic fracture energy measured through the pure-shear test remains unaffected by the number of vertical layers (i.e., number of unit cells along \( H \)), provided that it is sufficiently large.\cite{deng2023nonlocal} In a recent study, it was found that the number of unit cell layers required for the fracture energy of a rigid architected network (similar to monolithic architectures studied in this work) to converge is four.\cite{deng2023nonlocal} Beyond this threshold, the fracture energy remains unchanged with increasing sample dimensions, aligning with the Lake-Thomas model where all energy is released and dissipated on the layers near the crack tip.\cite{lake1967strength, deng2023nonlocal} Considering manufacturing constraints, we chose the sample dimensions in this work to be 16 unit cells along the width direction, 5 unit cells along the height direction, and 2 unit cells along the thickness direction. However, it is noteworthy that the number of layers required for the fracture energy to converge can differ based on the architecture. For highly nonlinear architectures, applicable to woven and DNI architectures studied here, the sample might necessitate more than 100 vertical layers for the intrinsic fracture energy value to stabilize (below which the fracture energy measured will be lower).\cite{deng2023nonlocal} Due to limitations of the fabrication technique used in this study, convergence studies to determine the number of layers needed for the fracture energy of the DNI metamaterial to stabilize were not feasible. Hence, the fracture energy measured in this work is presented as an underestimate of the actual fracture energy of the metamaterial with larger tessellations.

\newpage
\subsection{Supplementary Figures}
\renewcommand{\thefigure}{S\arabic{figure}}
\hfill\\
 \begin{figure}[H]
    \centering
    \includegraphics[width=\textwidth]{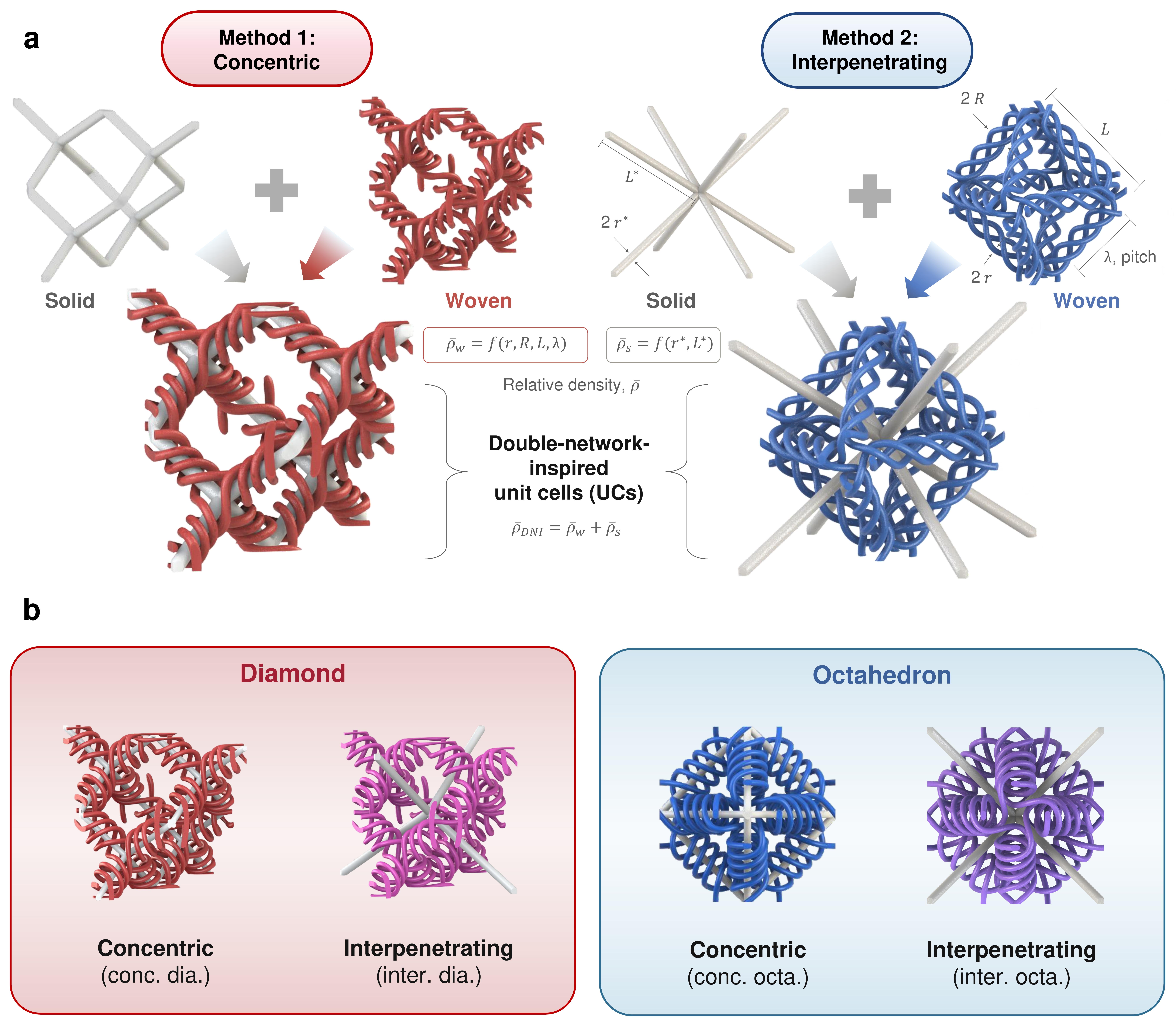}
    \caption{\textbf{Definition of design parameters and unit cell designs.} \textbf{a}, Illustration depicting the two design paradigms for creating double-network-inspired (DNI) woven metamaterials, (1) concentric and (2) interpenetrating designs, along with a set of design parameters defined for the creating the monolithic and woven networks. \textbf{b}, Different types of DNI unit cells investigated in this study: (i) concentric diamond, (ii) interpenetrating diamond, (iii) concentric octahedron, and (iv) interpenetrating octahedron.}
    \label{fig:designspace} 
\end{figure}

\newpage
 \begin{figure}[H]
    \centering
    \includegraphics[width=0.96\textwidth]{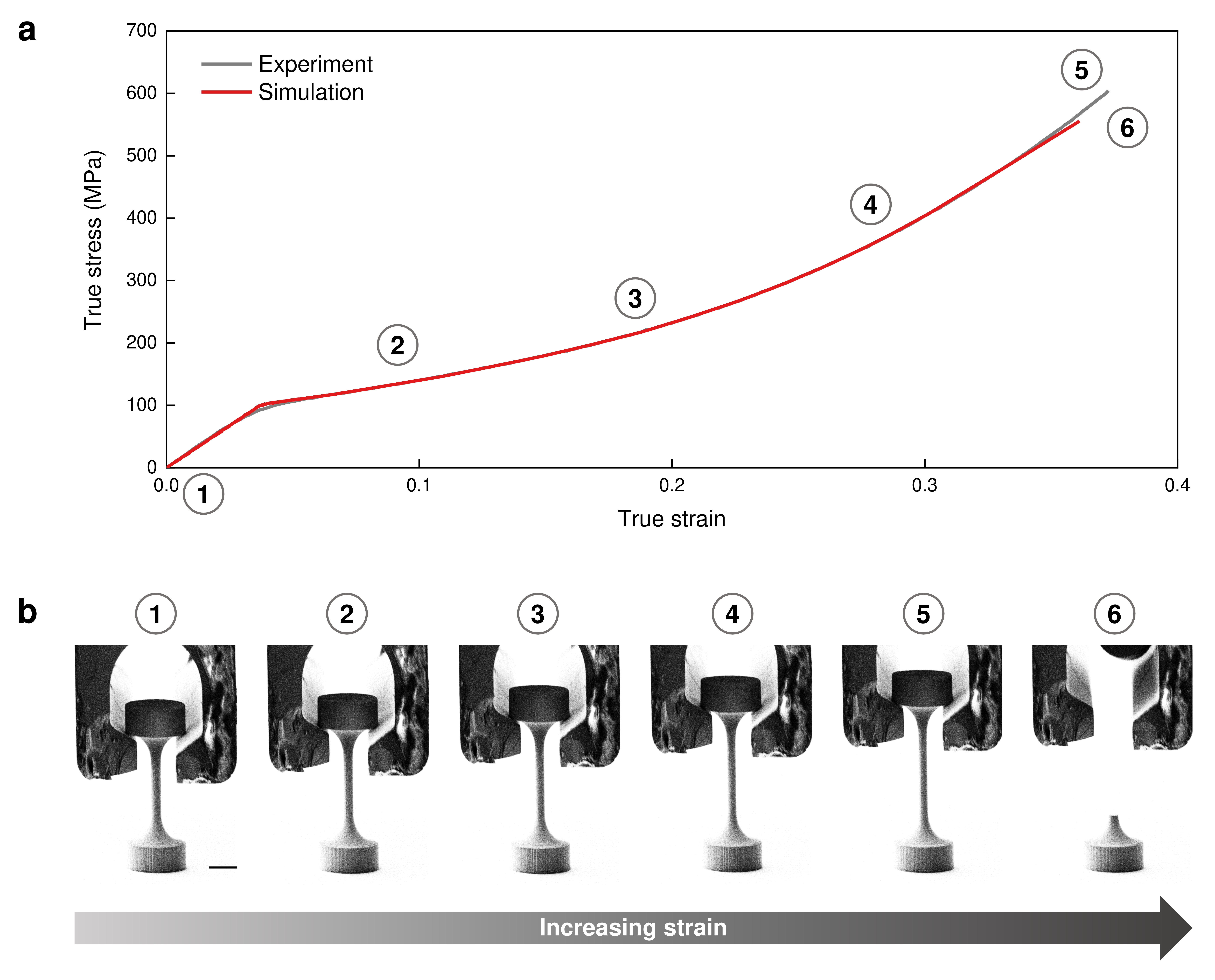}
    \caption{\textbf{\emph{In situ} tension of monolithic micropillars.} \textbf{a}, Comparison of the tensile true stress versus true strain curve for monolithic IP-Dip micropillars obtained via experiments and simulation. \textbf{b}, Snapshots showing the deformation behavior of the monolithic IP-Dip micropillars upon tension. Scale bar, \SI{10}{\micro\meter}.}
    \label{fig:pillar} 
\end{figure}

\newpage
 \begin{figure}[H]
    \centering
    \includegraphics[width=0.9\textwidth]{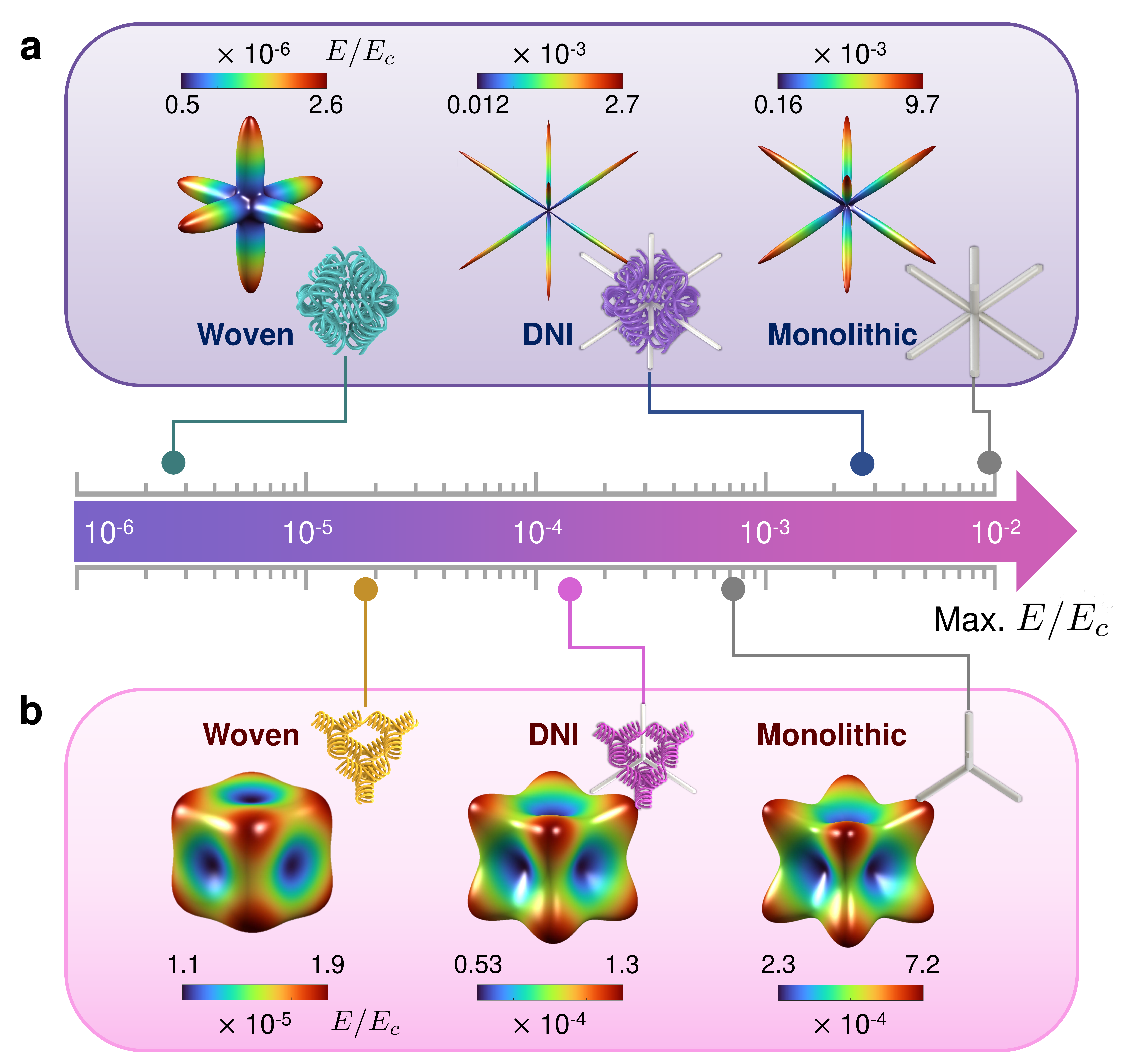}
     \caption{\textbf{Directional stiffness of interpenetrating DNI woven metamaterials.} The elastic surfaces of \textbf{a}, octahedron-based and \textbf{b}, diamond-based interpenetrating DNI architectures compared to their pure monolithic and woven counterparts at equivalent relative densities of 3.6$\pm$0.2\%, demonstrating the drastically enhanced stiffness of the DNI architectures compared to a pure woven architecture at the same mass. The Young's modulus of the architecture $E$ is normalized against the Young's modulus of the constituent material $E_{c}$. }
    \label{fig:elasticsurface2} 
\end{figure}

\newpage
 \begin{figure}[H]
    \centering
    \includegraphics[width=1\textwidth]{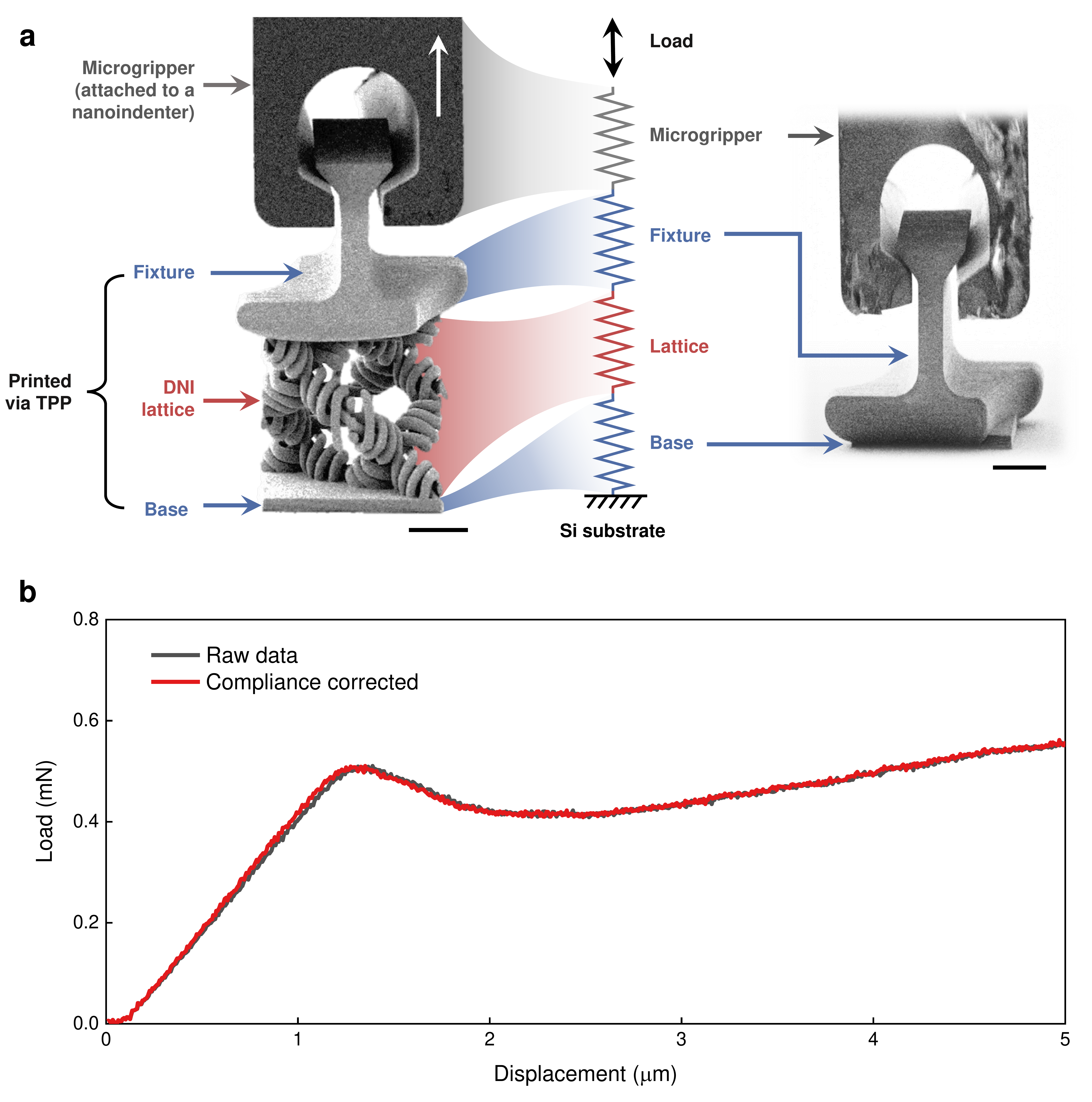}
    \caption{\textbf{Compliance correction.} \textbf{a}, \emph{In situ} tensile testing setup depicting the various components which make up the total compliance of the system, represented as springs in series. \textbf{b}, Load-displacement curves showing the difference between raw, uncorrected data with compliance corrected data for a DNI lattice unit cell.}
    \label{fig:compliance} 
\end{figure}

\newpage
 \begin{figure}[H]
    \centering
    \includegraphics[width=0.96\textwidth]{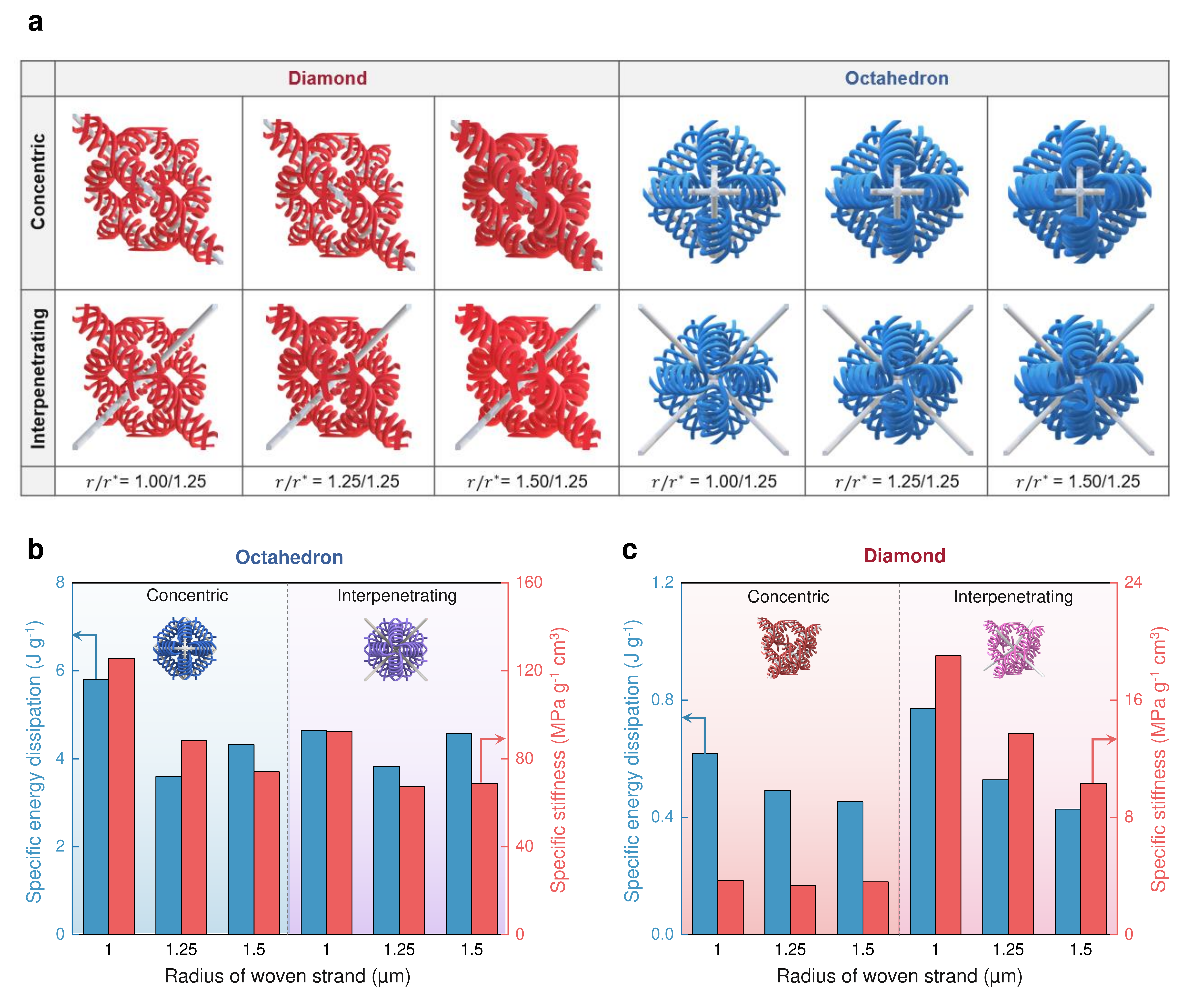}
    \caption{\textbf{Parametric study on DNI unit cells.} \textbf{a}, Illustrations showing the various types of DNI unit cells investigated in this work, in which the woven fiber radius $r$ is varied with respect to the monolithic strut radius $r*$. Comparison of the specific energy absorption and specific modulus for concentric and interpenetrating DNI unit cells based on \textbf{b}, octahedron and \textbf{c}, diamond designs with various woven-fiber radii.}
    \label{fig:optimization} 
\end{figure}

\newpage
 \begin{figure}[H]
    \centering
    \includegraphics[width=\textwidth]{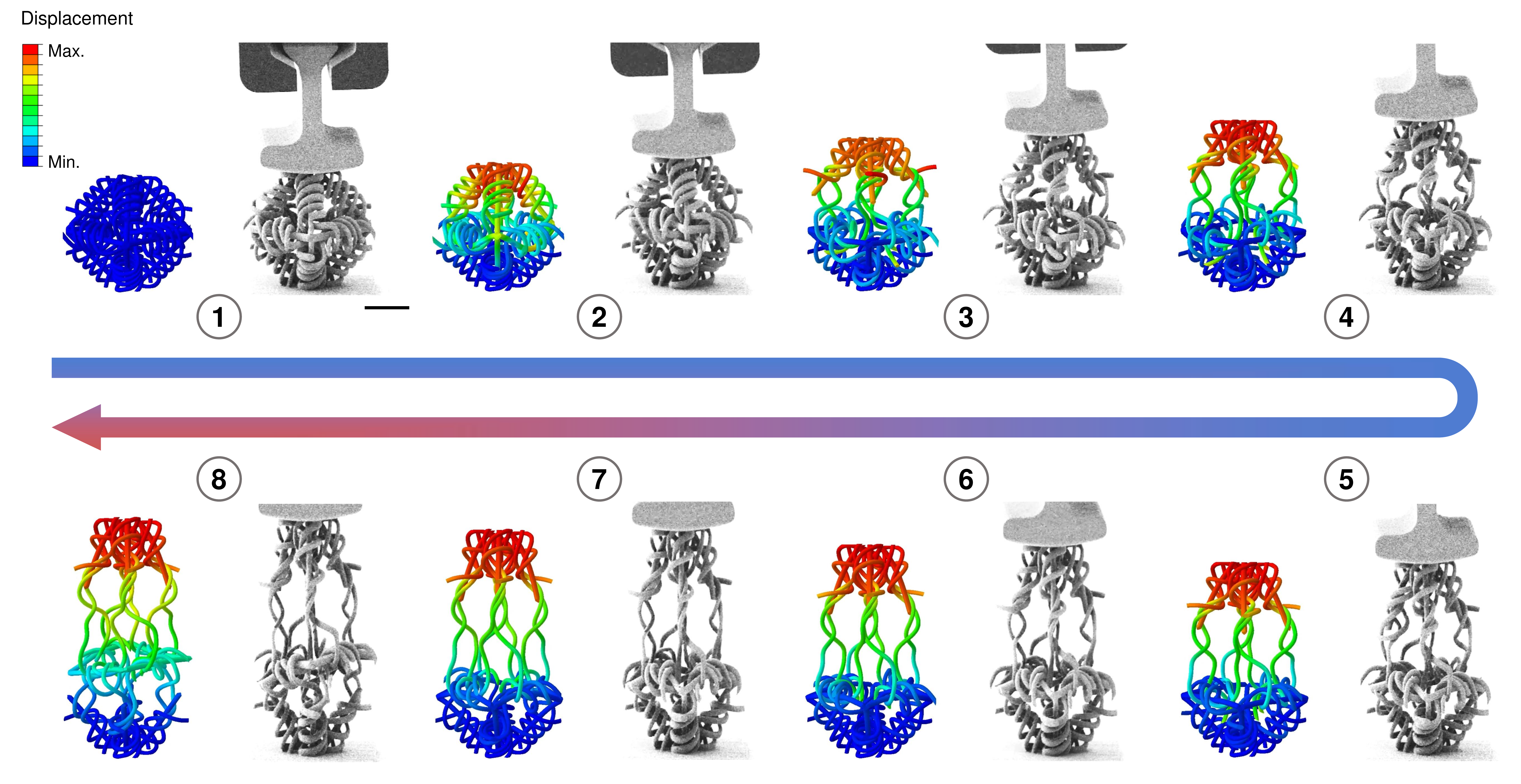}
    \caption{\textbf{Kinematic comparison of nonlinear finite element models (FEM) with \emph{in situ} experiments}. The validity of these computational models---with integrated frictional contact and cohesive surface interactions using a tetrahedral element representation---was confirmed by the visual agreement in the kinematics and failure of the architectures, when compared to the \emph{in situ} experiments under the same tensile boundary conditions. Specifically, we compare the deformation of woven fibers at each strain level (labeled from 1 to 8 in increasing order), the ensuing regions of entanglement and failure, and the strain at which failure occurs. Scale bar, 20 \textmu{}m.}
    \label{fig:simvsexp} 
\end{figure}

\newpage
 \begin{figure}[H]
    \centering
    \includegraphics[width=0.96\textwidth]{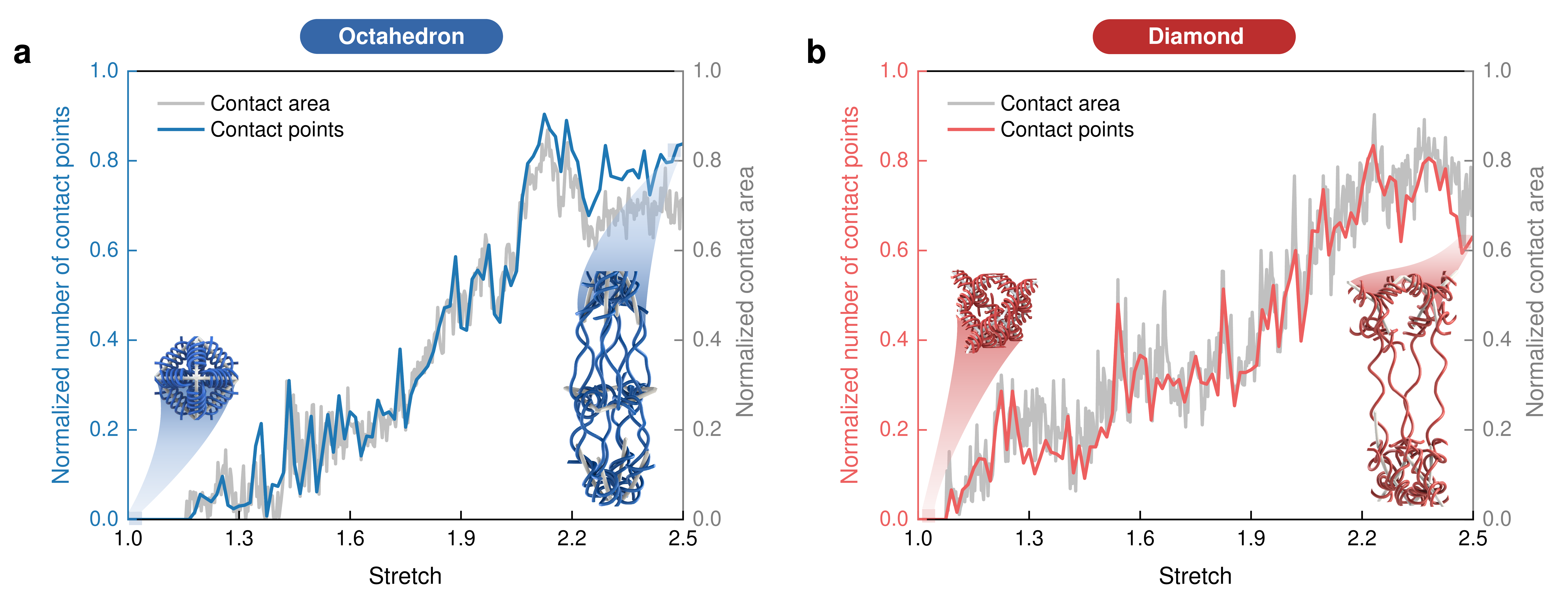}
    \caption{\textbf{Contact area approximation from contact points.} Nonlinear FEM with tetrahedral elements validating the close agreement between the change in contact area and contact points (number of paired elements in contact) upon tension for \textbf{a}, concentric octahedron and \textbf{b}, concentric diamond of up to a stretch of 2.5. Owing to the high computational cost of using tetrahedral elements for large tessellations of DNI metamaterials, we validate that contact points in beam-element simulations can be used to approximate contact area which is correlated to the degree of entanglements.}
    \label{fig:contactarea} 
\end{figure}

\newpage
 \begin{figure}[H]
    \centering
    \includegraphics[width=0.96\textwidth]{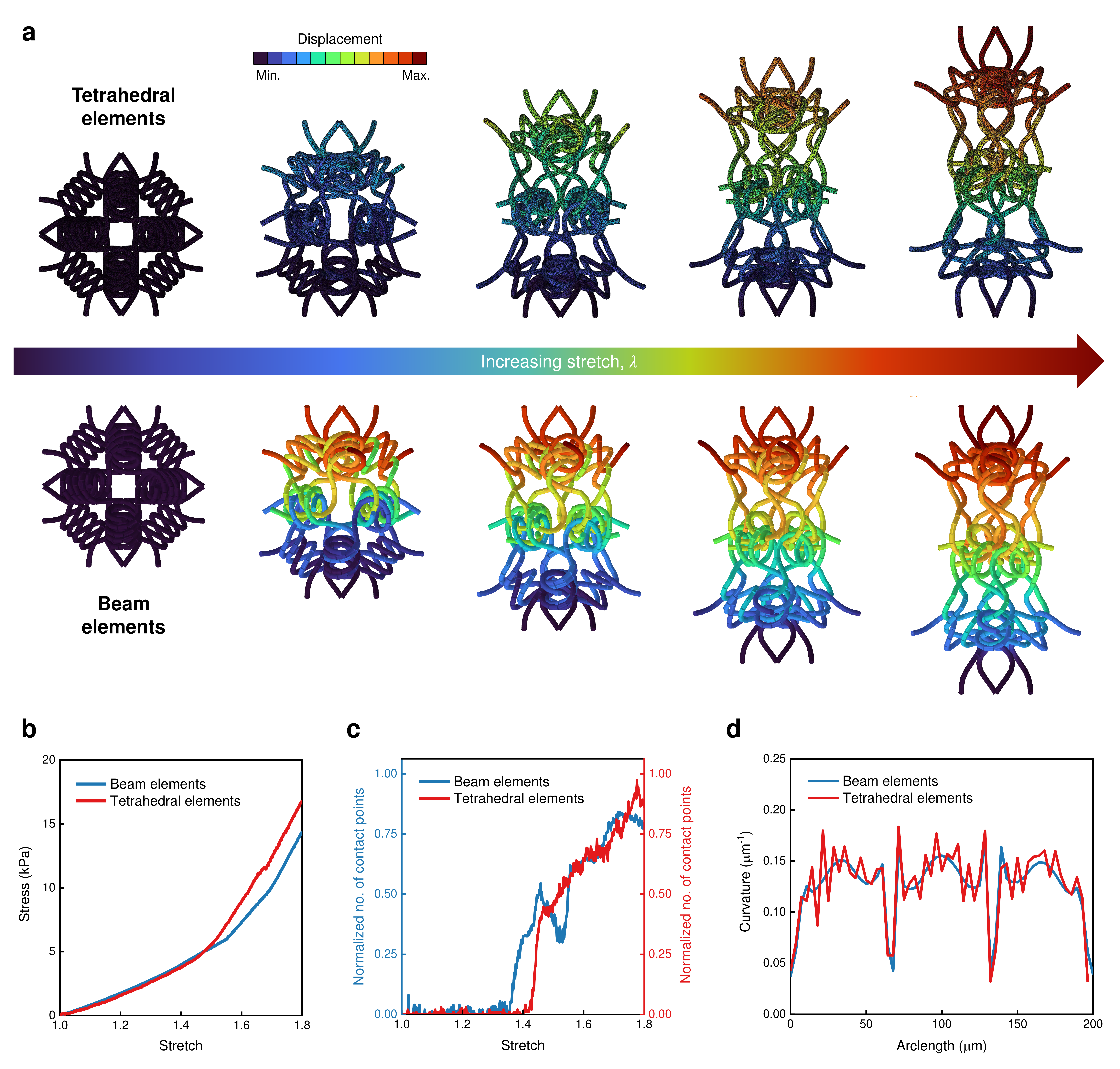}
    \caption{\textbf{Finite element modeling using tetrahedral elements versus beam elements.} \textbf{a}, Kinematic comparison between simulations conducted using tetrahedral elements and beam elements for a woven octahedron unit cell under tension. Comparison of results obtained between simulations performed using tetrahedral elements and beam elements in terms of their \textbf{b}, stress-stretch response, \textbf{c}, normalized contact points upon stretching, and \textbf{d}, curvature as a function of arclength for a representative woven fiber. }
    \label{fig:beamvs3d} 
\end{figure}

\newpage
 \begin{figure}[H]
    \centering
    \includegraphics[width=0.96\textwidth]{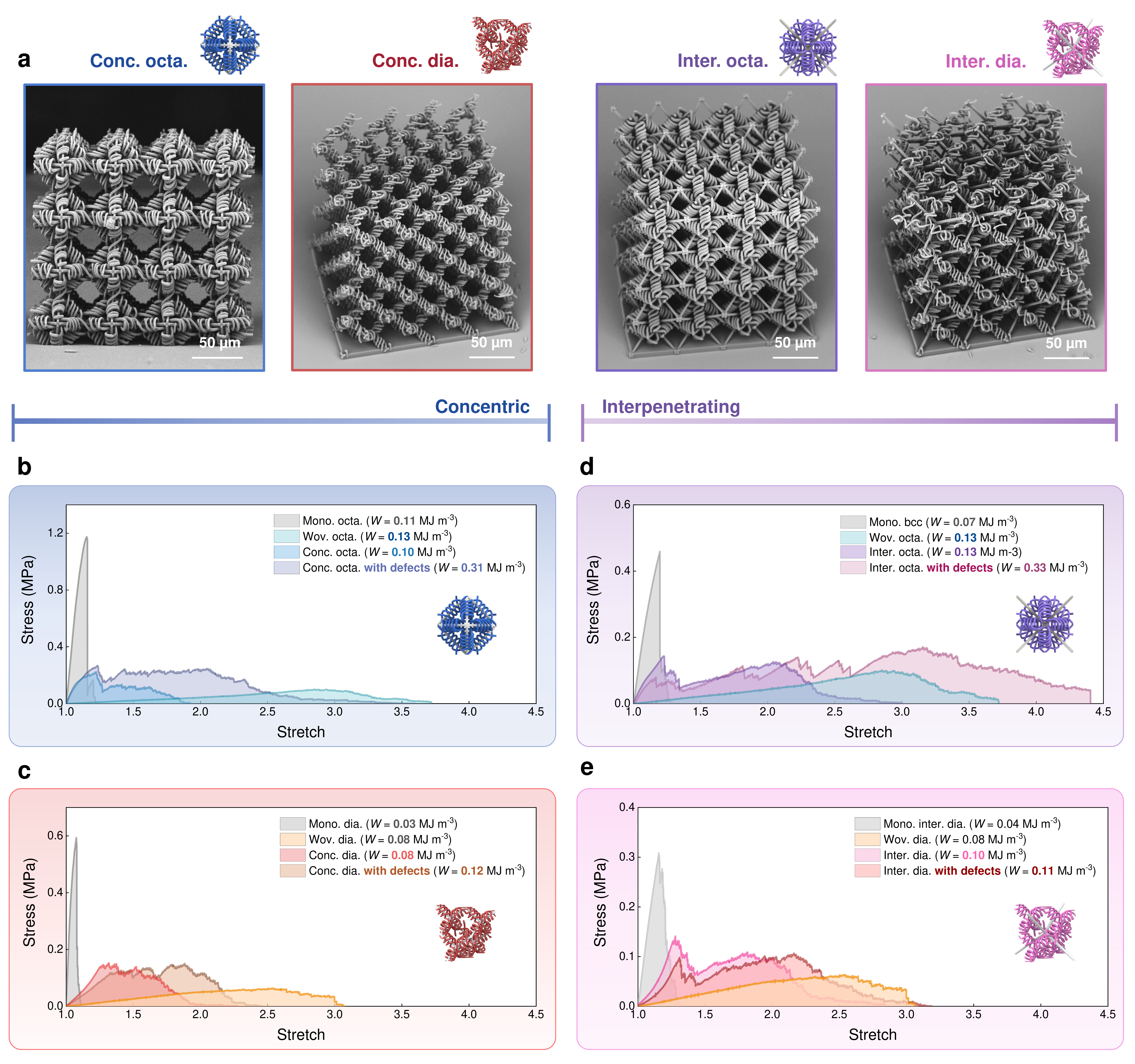}
    \caption{\textbf{Influence of incorporating internal defects (experiments).} \textbf{a}, SEM images of the four main types of DNI metamaterials studied: (i) concentric octahedron (conc. octa.), (ii) concentric diamond (conc. dia.), (iii) interpenetrating octahedron (inter. octa.), and (iv) interpenetrating diamond (inter. dia.). \textbf{b}--\textbf{e}, Comparison of the stress-stretch response for $4 \times 4 \times 4$ tessellations of DNI metamaterials (with and without internal defects) compared to pure monolithic and woven metamaterials at similar relative densities.}
    \label{fig:444exp} 
\end{figure}

\newpage
 \begin{figure}[H]
    \centering
    \includegraphics[width=0.96\textwidth]{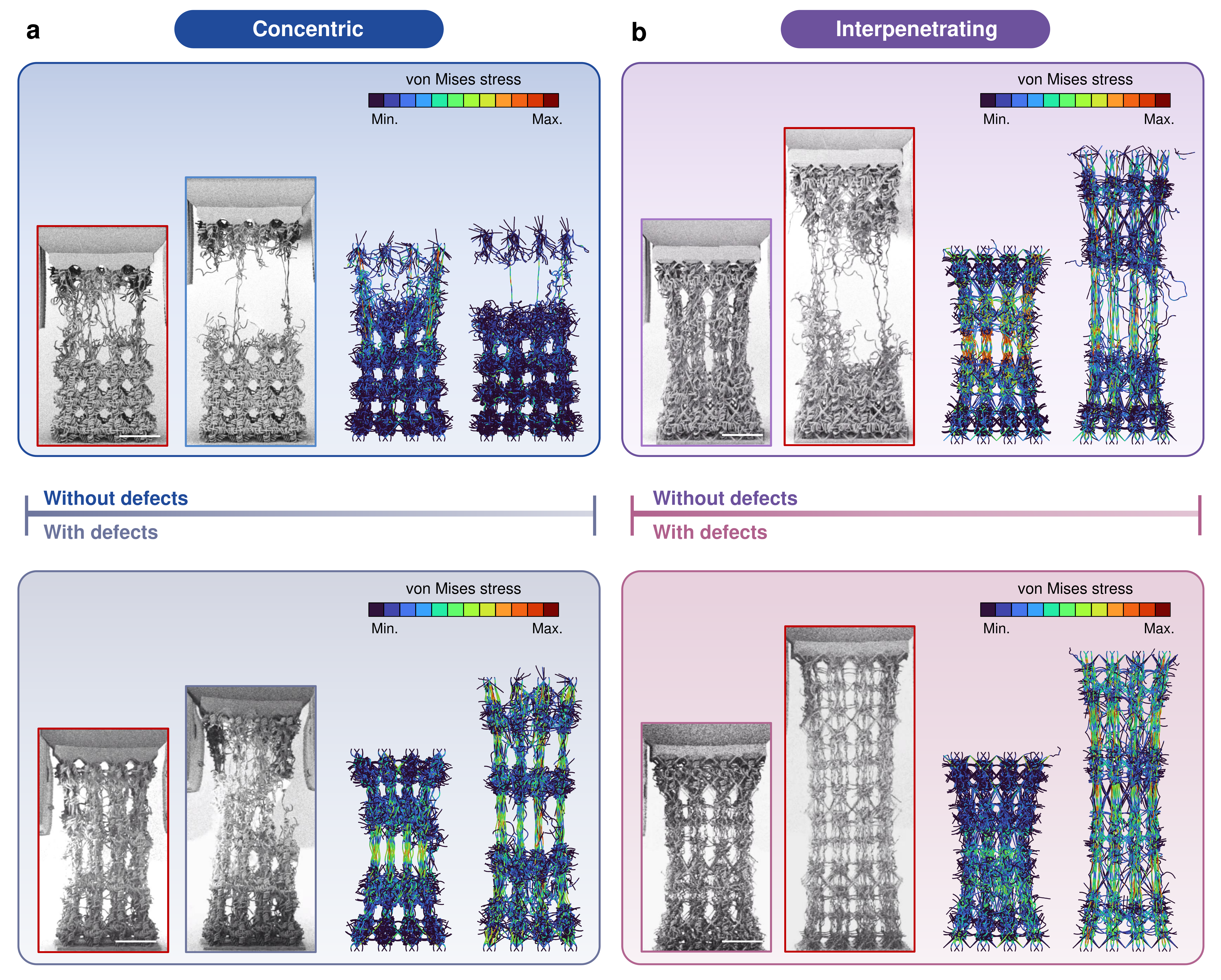}
    \caption{\textbf{Influence of incorporating internal defects (experiments versus simulations).} Comparison of simulation and experimental deformation behaviors of \textbf{a}, concentric and \textbf{b}, interpenetrating octahedron DNI metamaterials (with and without internal defects) upon uniaxial tension compared to the SEM snapshots taken during \emph{in situ} experiments. For the DNI metamaterials without incorporated internal defects, deformation is localized to the proximity of the fracture region of the monolithic network, resulting in reduced stretchability. For the DNI metamaterials with internal defects, a more uniform deformation of the woven network could be observed, resulting in enhanced stretchability and energy dissipation.}
    \label{fig:444sims} 
\end{figure}

\newpage
 \begin{figure}[H]
    \centering
    \includegraphics[width=0.96\textwidth]{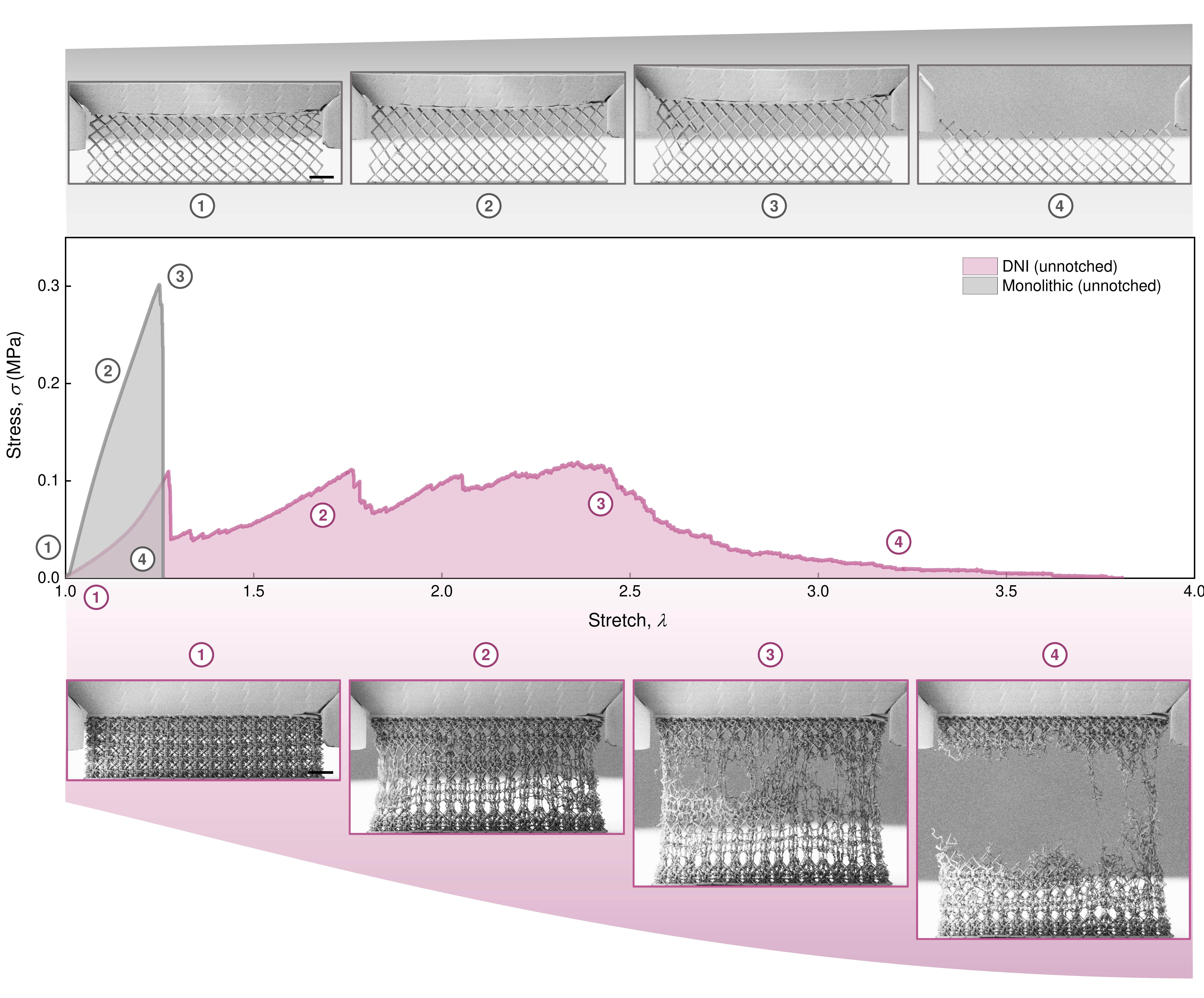}
    \caption{\textbf{Nominal stress-stretch curves of unnotched monolithic and DNI metamaterials along with SEM snapshots during tensile loading.} These experiments are used to determine the fracture energy of the monolithic and DNI metamaterials (when coupled with the results from notched samples). The SEM snapshots also confirm the drastically enhanced stretchability and energy dissipation performance of the DNI metamaterials compared to the monolithic metamaterials independent on the existence of a notch. Scale bars, \SI{100}{\micro\meter}.}
    \label{fig:exppureshear} 
\end{figure}

\newpage
 \begin{figure}[H]
    \centering
    \includegraphics[width=0.96\textwidth]{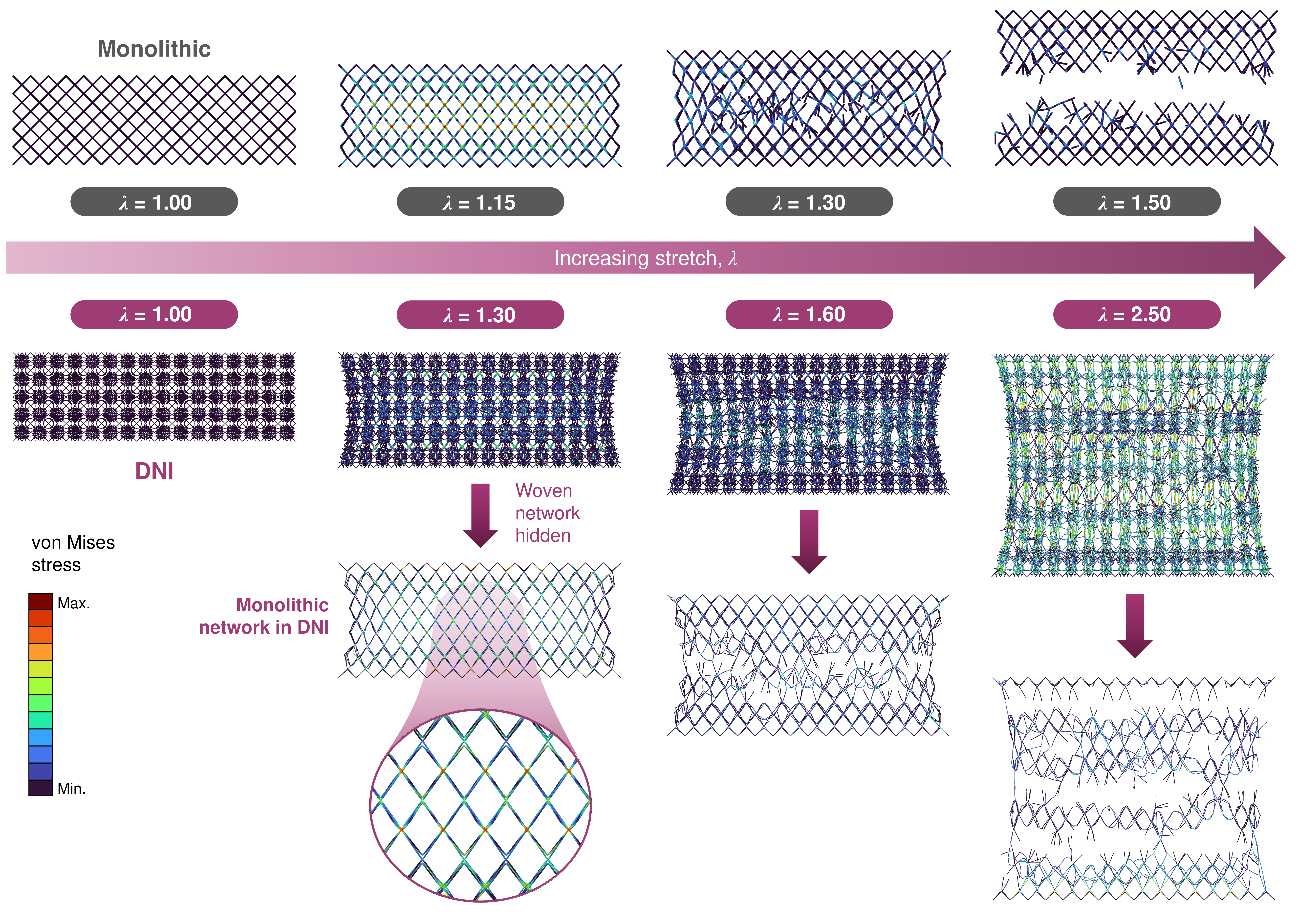}
    \caption{\textbf{Simulated tensile response of unnotched monolithic and DNI metamaterial.} In the deformed morphology of the notched DNI metamaterial, the isolated monolithic network in the deformed DNI morphology shows a large damage zone in which failure occurs at multiple layers. In the simulated fractured network of the pure monolithic metamaterial, failure occurs only in the vicinity of the initial damage region.}
    \label{fig:simspureshear} 
\end{figure}

\newpage
\subsection{Supplementary videos \& captions}
\hfill

\noindent\textbf{Supplementary video 1.} \\
\hspace{1em}\emph{In situ} tension of monolithic IP-Dip micropillars synched with corresponding engineering stress-strain curve. \href{https://www.dropbox.com/scl/fi/iomaidi1wqj4eonw87py8/Supplementary-Video-1-v2.mp4?rlkey=8b9vvpep36pgzdvzjpaxr032o&dl=0}{(link)}

\noindent\textbf{Supplementary video 2.} \\
\emph{In situ} tension of monolithic, woven, and double-network-inspired (DNI) unit cells based on concentric octahedron. \href{https://www.dropbox.com/scl/fi/kcclfg0wb0mrovhwn3dcq/Supplementary-Video-2.mp4?rlkey=blonlmzvuewyomg5egyap2fv3&dl=0}{(link)}

\noindent\textbf{Supplementary video 3.} \\
\emph{In situ} tension of monolithic, woven, and double-network-inspired (DNI) unit cells based on concentric diamond. \href{https://www.dropbox.com/scl/fi/re1bcc89ps0ufb9sptii8/Supplementary-Video-3.mp4?rlkey=46w2bhq9lpy7s1qxps4pr26uk&dl=0}{(link)}

\noindent\textbf{Supplementary video 4.} \\
\emph{In situ} tension of octahedron woven unit cells with various woven fiber radii.
\href{https://www.dropbox.com/scl/fi/q8t0xpsbyqosr2uo69r7h/Supplementary-Video-4.mp4?rlkey=o2q0esk8u6csdj4xq87pagal8&dl=0}{(link)}

\noindent\textbf{Supplementary video 5.} \\
\emph{In situ} tension of diamond woven unit cells with various woven fiber radii. \href{https://www.dropbox.com/scl/fi/ed56g5r9mal3ft3i8osk5/Supplementary-Video-5.mp4?rlkey=xdqox0mpwjmd83yhtael7n51l&dl=0}{(link)}

\noindent\textbf{Supplementary video 6.} \\
\emph{In situ} tension of  double-network-inspired (DNI) unit cells based on concentric octahedron for various woven fiber radii. 
\href{https://www.dropbox.com/scl/fi/ojqgxszp1k9bdd24gjfhh/Supplementary-Video-6.mp4?rlkey=u0r1k2thxmpa43r2qrlgmo7k0&dl=0}{(link)}

\noindent\textbf{Supplementary video 7.} \\
\emph{In situ} tension of  double-network-inspired (DNI) unit cells based on concentric diamond for various woven fiber radii. 
\href{https://www.dropbox.com/scl/fi/393x0uj3mk986s38bd2pv/Supplementary-Video-7.mp4?rlkey=9iglltc4f1v4aze1yez4vrfq6&dl=0}{(link)}

\noindent\textbf{Supplementary video 8.} \\
Comparison of the simulated deformation behaviors of concentric octahedron unit cells with tetrahedral elements against the experimental results. 
\href{https://www.dropbox.com/scl/fi/swa7nlk54z9kjanwuaq61/Supplementary-Video-8.mp4?rlkey=unhlh8aytb97vk48647h1s0jp&dl=0}{(link)}

\noindent\textbf{Supplementary video 9.} \\
Comparison of the simulated deformation behaviors of concentric octahedron unit cells meshed with beam elements against those obtained via meshing with tetrahedral elements. 
\href{https://www.dropbox.com/scl/fi/673uz1sl49hj391psjr97/Supplementary-Video-9.mp4?rlkey=qmdiofuv00pyr1d9qmkj7z4t6&dl=0}{(link)}

\noindent\textbf{Supplementary video 10.} \\
\emph{In situ} tension of $4 \times 4 \times 4$ double-network-inspired (DNI) metamaterials based on concentric octahedron (with and without internal void defects).
\href{https://www.dropbox.com/scl/fi/2km79thl57rlhvvj14nix/Supplementary-Video-10v2.mp4?rlkey=s6pgdbv1h90hz0lpy8xx5ea6f&dl=0}{(link)}

\newpage
\noindent\textbf{Supplementary video 11.} \\
\emph{In situ} tension of $4 \times 4 \times 4$ double-network-inspired (DNI) metamaterials based on interpenetrating octahedron (with and without internal void defects).
\href{https://www.dropbox.com/scl/fi/ksty9ea2r83dgl6qsbgg7/Supplementary-Video-11.mp4?rlkey=vt54m8pmfuha6af3fewqtjwgu&dl=0}{(link)}

\noindent\textbf{Supplementary video 12.} \\
Pure-shear experiments on notched monolithic and double-network-inspired (DNI) metamaterial samples to determine the critical stretch at which the crack starts to propagate globally.
\href{https://www.dropbox.com/scl/fi/1j9ljuuw9yc1gavdcl4xq/Supplementary-Video-12.mp4?rlkey=7118c75wyl2yyftu68c3d43qb&dl=0}{(link)}

\noindent\textbf{Supplementary video 13.} \\
Beam element simulations of pure-shear experiments on notched monolithic and double-network-inspired (DNI) metamaterials
Pure-shear experiments on unnotched monolithic and double-network-inspired (DNI) metamaterials.
\href{https://www.dropbox.com/scl/fi/8ncm5qx8ql9au02o9ukm6/Supplementary-Video-13.mp4?rlkey=3wopbna3ba123908esrnvspl8&dl=0}{(link)}

\noindent\textbf{Supplementary video 14.} \\
Pure-shear experiments on unnotched monolithic and double-network-inspired (DNI) metamaterials.
\href{https://www.dropbox.com/scl/fi/zvo2acfxemoj3jzwkfypz/Supplementary-Video-14.mp4?rlkey=e2kclfr40tpqx1zreq2t4alm4&dl=0}{(link)}

\noindent\textbf{Supplementary video 15.} \\
Beam element simulations of pure-shear experiments on unnotched monolithic and double-network-inspired (DNI) metamaterials.
\href{https://www.dropbox.com/scl/fi/bkbo3gozgt42rcyw0lhx0/Supplementary-Video-15.mp4?rlkey=i18k97piqkiulop696tlz5uw3&dl=0}{(link)}


\nolinenumbers
\vspace{10pt}

\newpage
\subsection{References}
{\spacing{1}
\bibliography{references}
}
\bibliographystyle{naturemag.bst}